\pdfoutput=1
\documentclass{JFM-FLM_Au}
\usepackage{subcaption}
\usepackage{graphicx}
\usepackage[version=4]{mhchem}
\usepackage{amssymb}
\usepackage{xcolor}

\lefttitle{M. Fratini, P.S. Volpiani, and M. Bernardini}
\righttitle{Journal of Fluid Mechanics}

\title{Effects of thermochemical modelling on a hypersonic shock-wave/turbulent boundary-layer interaction}

\author{Marco Fratini\aff{1}, Pedro Stefanin Volpiani\aff{2} \and Matteo Bernardini\aff{1}}

\affiliation{\aff{1} Department of Mechanical and Aerospace Engineering, Sapienza University of Rome, via Eudossiana 18,
00184 Rome, Italy \aff{2} ONERA, The French Aerospace Lab, 8 Rue des Vertugadins, 92190 Meudon, France}

\corresau{Matteo Bernardini, \email{matteo.bernardini@uniroma1.it}}

\begin{document}

\maketitle
\begin{abstract}
Thermochemical non-equilibrium can alter the structure, loads, and time scales of hypersonic shock-wave/turbulent boundary-layer interactions, yet its role in fully turbulent configurations remains largely unquantified. The present work addresses this issue by performing three direct numerical simulations of an oblique shock impinging on a turbulent high-enthalpy boundary layer at edge Mach number $M_e=6.4$ and stagnation enthalpy $H_e=16.9$~MJ/kg. The simulations share identical geometry and freestream conditions, but employ a hierarchy of progressively simplified thermochemical descriptions: a finite-rate reactive case, a single-species thermally perfect gas model, and a single-species calorically perfect model. The reactive simulation shows that the shock-induced temperature rise substantially enhances chemical activity relative to the incoming boundary layer, with peak concentrations of dissociation products attained downstream of the interaction. Thus, the thermal and chemical responses are not synchronised: the composition lags the rapid thermal forcing imposed by the shock system, and turbulent Damköhler numbers reach values of order unity within the recirculation region, indicating non-negligible turbulence–chemistry interaction. The comparison among the three models shows that thermally and calorically perfect descriptions yield similar predictions, whereas finite-rate chemistry produces systematic differences: a smaller separation bubble, lower post-interaction wall heat flux, lower mean and fluctuating temperatures, and a less inclined reflected shock. In the present regime, the dominant modelling distinction is therefore between frozen and chemically reacting descriptions, with caloric-model effects playing only a secondary role. We conclude that simplified fluid models provide conservative estimates of wall heat flux and separation extent, making them well suited for hypersonic thermal protection system design.

\end{abstract}

\section{Introduction}
Shock-wave/boundary-layer interaction (SBLI) is a longstanding cornerstone problem of high-speed aerodynamics because it strongly affects wall-pressure loads, skin friction, wall-heat flux, boundary-layer separation, and flow unsteadiness in a wide range of aerospace applications, including control surfaces, engine intakes, over-expanded nozzles, and re-entry vehicles \citep{dolling2001fifty, gaitonde2015progress}. Despite decades of research, SBLI remains difficult to predict accurately because the shock system, the incoming boundary layer, and the separated flow region interact over a broad range of spatial and temporal scales. In particular, one of the most debated aspects of turbulent SBLI is the low-frequency unsteadiness of the interaction, often observed as a broadband streamwise motion of the reflected shock and of the associated separated region. 

For hypersonic flows, the problem becomes even richer. The stronger kinetic-to-internal energy conversion produced by viscous dissipation may activate additional physical processes, including vibrational excitation, molecular dissociation, and, at more extreme conditions, ionisation. As emphasised by \cite{candler2019rate}, hypersonic flows are governed by the competition among multiple processes whose characteristic times may become comparable with the fluid dynamic timescale. As a result, the gas state may depart from local thermodynamic equilibrium, and the resulting rate effects can alter the overall dynamics. Although these issues are widely recognised in hypersonic aero-thermodynamics, the large majority of studies of turbulent SBLI are still carried out under calorically perfect gas assumption, or at most with thermally perfect models.

Even in the absence of thermochemical non-equilibrium, SBLI has already been shown to be highly sensitive to the wall thermal condition. Numerical studies in both supersonic and hypersonic cryogenic regimes have demonstrated that wall cooling can significantly affect the interaction length, the onset and extent of separation, wall heat transfer, and the pressure fluctuation levels \citep{volpiani2018effects,volpiani2020effects,zhang2024effects}. 
In particular, direct numerical simulations (DNS) of low-enthalpy hypersonic turbulent SBLI by \citet{volpiani2020effects} showed that wall cooling tends to mitigate flow separation while modifying wall-pressure fluctuations and thermal loads, and also highlighted that a major part of this sensitivity is mediated by changes in the state of the incoming boundary layer. This observation is especially relevant in the hypersonic regime, where strong wall cooling is often unavoidable in practice and where the incoming boundary layer may itself be profoundly altered by thermodynamic effects. Moreover, recent DNS of hypersonic turbulent compression-corner flow by \citet{direnzo2024stagnation} showed that replacing a calorically perfect-gas model with a calorically imperfect description including vibrational excitation does not qualitatively alter the flow organisation, but can still induce non-negligible quantitative variations in skin friction, wall heat flux, and wall-pressure fluctuations.

At the same time, a growing body of work has shown that thermochemical non-equilibrium can substantially modify hypersonic turbulent boundary layers even in the absence of shock impingement. Following early studies on temporally evolving, high-temperature and chemically reactive boundary layers \citep{martin2001temperature,duan2009effect,duan2011direct4}, more recent direct numerical simulations have documented the role of vibrational non-equilibrium and finite-rate chemistry in both transitional and turbulent high-enthalpy, spatially evolving, boundary layers \citep{urzay2021direct,passiatore2022thermochemical,williams2025turbulence,fratini2026wall}. Taken together, these works indicate that, under sufficiently energetic conditions, high-enthalpy effects are not a secondary correction, but can exert a first-order influence on the prediction of both mechanical and thermal loads. This also implies that the use of simplified frozen models in hypersonic wall-bounded flows must be assessed carefully, rather than assumed a priori to be adequate.

By contrast, the literature on hypersonic high-enthalpy SBLI remains comparatively sparse \citep{raje2025}. The interaction between a turbulent boundary layer and an impinging shock sharply reduces the viscous length scale downstream of the interaction, thereby making the resolution requirements significantly more severe than in canonical zero-pressure-gradient boundary layers. As a result, direct numerical simulations of hypersonic high-enthalpy flows have so far focused mainly on boundary-layer configurations, whereas SBLI cases remain much less explored.
A first step in this direction was taken by \citet{volpiani2021numerical}, who presented an early three-dimensional turbulent SBLI including chemical non-equilibrium. That numerical simulation, however, was intentionally carried out at low Reynolds number ($Re_\tau \approx 200$) in order to keep the problem computationally tractable, and therefore its main contribution lies in demonstrating the feasibility of the approach rather than in providing a fully developed turbulent reference case.
More recently, \citet{passiatore2023shock} investigated the impingement of an oblique shock on a transitional high-enthalpy boundary layer out of thermochemical equilibrium. Their results showed that thermal non-equilibrium is enhanced and sustained by the shock-induced transition, whereas chemical activity remains comparatively weak. These studies clearly demonstrate both the feasibility and the relevance of high-fidelity simulations of high-enthalpy SBLI, but also highlight how limited the currently available database still is. To the best of the authors' knowledge, a direct numerical simulation study of a chemically reacting hypersonic SBLI under fully turbulent high-Reynolds-number conditions is still missing. To address this gap, the multispecies reactive Navier–Stokes equations were solved using a high-fidelity DNS approach on meshes exceeding $10^{10}$ grid points, leveraging GPU-accelerated computing to achieve statistically converged results over extended simulation times.

The present work addresses an additional question concerning the role of fluid modelling in turbulent hypersonic SBLI. This is investigated through three direct numerical simulations performed under identical geometry, external flow conditions, and numerical framework, while employing progressively simplified thermochemical descriptions. The reference case is a fully reactive simulation accounting for finite-rate chemistry. Two additional simplified cases are then considered: a thermally perfect frozen model and a calorically perfect frozen model. This hierarchy makes it possible to separate the contribution of finite-rate chemistry from that of the thermal inertia and to quantify how the fluid model affects the incoming boundary layer, the interaction topology, the thermal and chemical response, and the wall loads. Besides clarifying the physics of thermochemical non-equilibrium in a hypersonic SBLI, this comparison also helps assess whether simplified frozen models remain adequate for engineering predictions.

The paper is organised as follows. \S~\ref{sec:methodology} describes the governing equations and the numerical methods employed; \S~\ref{sec:casedescription} presents the numerical setup of the simulations; \S~\ref{sec:blvalidation} validates the incoming boundary layer; \S~\ref{sec:reactivesimulation} presents the results of the reactive simulation; \S~\ref{sec:fluidmodeling} discusses the effects of fluid modelling on the SBLI; \S~\ref{sec:conclusions} summarises the main conclusions.

\section{Methodology}
\label{sec:methodology}

\subsection{Governing equations}
\label{subsec:governingeq}

The simulations solve the compressible Navier--Stokes equations for a chemically reacting mixture of $N_s$ thermally perfect gases, under the assumption of thermal equilibrium among vibrational modes of the molecules. For species, momentum, and total energy, the conservation equations read

\begin{equation}
\label{eq:masscons}
\frac{\partial (\rho Y_n)}{\partial t}+\frac{\partial (\rho Y_n u_j)}{\partial x_j}+\frac{\partial (\rho Y_n u^D_{n,j})}{\partial x_j}=\dot{\omega}_n,\qquad n=1,\dots,N_s
\end{equation}
\begin{equation}
\label{eq:momcons}
\frac{\partial (\rho u_i)}{\partial t}+\frac{\partial (\rho u_i u_j)}{\partial x_j}+\frac{\partial (p\delta_{ij}-\sigma_{ij})}{\partial x_j}=0,\qquad i=1,2,3
\end{equation}
\begin{equation}
\label{eq:encons}
\frac{\partial (\rho E)}{\partial t}+\frac{\partial (\rho H u_j)}{\partial x_j}+\frac{\partial q_j}{\partial x_j}-\frac{\partial (\sigma_{ij}u_i)}{\partial x_j}+\frac{\partial}{\partial x_j}\left(\rho\sum_{n=1}^{N_s}Y_n h_n u^D_{n,j}\right)=0
\end{equation}

Here $x$, $y$, and $z$ denote the streamwise, wall-normal, and spanwise directions, respectively; $u_i$ are the velocity components, $\rho$ is the density, $p$ is the pressure, $T$ is the temperature, and $Y_n$ is the mass fraction of species $n$.

Species diffusion is modelled with a Fickian approximation corrected to enforce zero net diffusive mass flux,

\begin{equation}
 u^D_{n,j}=-\frac{D_n}{X_n}\frac{\partial X_n}{\partial x_j}+\sum_{m=1}^{N_s}\left(D_m\frac{Y_m}{X_m}\frac{\partial X_m}{\partial x_j}\right)
\end{equation}

where $X_n$ is the molar fraction and $D_n$ is the mixture diffusivity of species $n$, obtained from binary diffusivities through the mixing rule of \citet{bird2006transport},

\begin{equation}
D_n=\frac{1-Y_n}{\sum_{\substack{m=1 \\ m\neq n}}^{N_s} X_m/D_{nm}}
\end{equation}

The viscous stress tensor is

\begin{equation}
\sigma_{ij}=\mu\left(\frac{\partial u_i}{\partial x_j}+\frac{\partial u_j}{\partial x_i}-\frac{2}{3}\frac{\partial u_k}{\partial x_k}\delta_{ij}\right)
\end{equation}

where the mixture viscosity $\mu$ is computed from the species viscosities $\mu_n$ using Wilke's rule \citep{wilke1950viscosity},

\begin{equation}
\mu=\sum_{n=1}^{N_s}\frac{Y_n\mu_n}{\sum_{m=1}^{N_s}G_{mn}(\frac{\mathcal{M}_n}{\mathcal{M}_m})Y_m},\quad
G_{nm}=\frac{1}{\sqrt{8}}\left(1+\frac{\mathcal{M}_n}{\mathcal{M}_m}\right)^{-1/2}\left[1+\left(\frac{\mu_n}{\mu_m}\right)^{1/2}\left(\frac{\mathcal{M}_n}{\mathcal{M}_m}\right)^{1/4}\right]^2
\end{equation}

Total energy is written as in \cite{poinsot2005theoretical}

\begin{equation}
E=H-\frac{p}{\rho}=\sum_{n=1}^{N_s}Y_n h_n-\frac{p}{\rho}+\frac{1}{2}u_i u_i
\end{equation}

with species enthalpy

\begin{equation}
h_n=\Delta h_{f,n}^{T_{\mathrm{ref}}}+\int_{T_{\mathrm{ref}}}^{T} c_{p,n}(T')\,\mathrm{d}T'
\end{equation}

where $T_{\mathrm{ref}}=298.15\,\mathrm{K}$. Thermodynamic properties are evaluated via NASA polynomials \citep{mcbride2002nasa}.

The conductive heat flux is modelled as

\begin{equation}
q_j=-\lambda\frac{\partial T}{\partial x_j}
\end{equation}

where $\lambda$ is obtained from species conductivities $\lambda_n$ using the Mathur mixing rule \citep{mathur1967thermal},

\begin{equation}
\lambda=\frac{1}{2}\left[\sum_{n=1}^{N_s}X_n\lambda_n+\left(\sum_{n=1}^{N_s}\frac{X_n}{\lambda_n}\right)^{-1}\right].
\end{equation}

The system is closed with the ideal-gas equation of state

\begin{equation}
\label{eq:eos}
p=\rho R T,
\end{equation}
where $R=\sum_{n=1}^{N_s} Y_n\mathcal{R}/\mathcal{M}_n$, and $\mathcal{R}$ is the universal gas constant.

High-temperature air is modelled with the 5-species, 5-reaction Park mechanism \citep{park1990nonequilibrium}, valid in the present temperature range where ionisation effects remain negligible \citep{urzay2021direct}:

\begin{align}
\ce{N2 + M &<=> 2N + M}    \tag{R1}\label{eq:R1}\\
\ce{O2 + M &<=> 2O + M}    \tag{R2}\label{eq:R2}\\
\ce{NO + M &<=> N + O + M} \tag{R3}\label{eq:R3}\\
\ce{N2 + O &<=> NO + N}    \tag{R4}\label{eq:R4}\\
\ce{NO + O &<=> O2 + N}    \tag{R5}\label{eq:R5}
\end{align}

Reactions \eqref{eq:R1}--\eqref{eq:R3} are dissociation steps, whereas \eqref{eq:R4}--\eqref{eq:R5} are exchange reactions. The chemical source term in Eq.~\eqref{eq:masscons} is computed from the law of mass action,

\begin{align}
\label{eq:wdot}
\dot{\omega}_n=\mathcal{M}_n\sum_{r=1}^{N_r}(\nu''_{nr}-\nu'_{nr}) \sum_{s=1}^{N_s} F_{sr} \left(\frac{\rho Y_s}{\mathcal{M}_s}\right) \left[k_{f,r}\prod_{m=1}^{N_s}\left(\frac{\rho Y_m}{\mathcal{M}_m}\right)^{\nu'_{mr}}-k_{b,r}\prod_{m=1}^{N_s}\left(\frac{\rho Y_m}{\mathcal{M}_m}\right)^{\nu''_{mr}}\right]
\end{align}
in which $\nu_{nr}'$ and $\nu_{nr}''$ are the stoichiometric coefficients of the $n-th$ species in the $r-th$ reaction on the reactants and products sides, respectively. $F_{nr}$ are the third-body efficiencies, $k_{f,r}$ are the forward rates computed using the Arrhenius law and $k_{b,r}$ are the backward rates, evaluated as $k_{b,r}=k_{f,r}/K_{eq,r}$, with $K_{eq,r}$ being the equilibrium constants.

\subsection{Numerical methods}
\label{subsec:nummethods}

The governing equations are solved on a Cartesian grid using an extended version of the Supersonic TuRbulEnt Accelerated Navier-Stokes Solver (STREAmS) 2.0 \citep{bernardini2021streams,bernardini2023streams2}. Convective terms are discretised with a hybrid strategy: a sixth-order central energy-preserving scheme in smooth regions, and a fifth-order Weighted Essentially Non-Oscillatory (WENO) reconstruction with Lax--Friedrichs flux splitting near discontinuities. The hybridisation between the central and the WENO schemes is employed via the use of the Ducros' shock sensor \citep{ducros1999large}. Viscous fluxes are discretised with a second-order conservative central scheme, and time advancement is performed with a third-order, three-stage Runge--Kutta method.

The solver has been ported to modern GPU platforms (NVIDIA, AMD, and Intel) and exhibits excellent scalability \citep{sathyanarayana2025high}. For multicomponent thermochemical simulations, temperature-dependent species properties ($\mu_n(T),\:\lambda_n(T),\:D_{nm}(T)$) and equilibrium constants $K_{eq,r}(T)$ are tabulated over a wide range of temperatures during preprocessing and transferred to device memory; mixture properties and reaction rates are then evaluated at runtime. The implementation has been validated against canonical reacting-flow benchmarks \citep{ferrer2014detailed} and in dedicated verification/validation campaigns \citep{forti2025development}.

In the following, the overbar $(\overline{f})$ denotes Reynolds averaging,
whereas the tilde $(\widetilde{f})$ denotes Favre averaging,
defined as $\widetilde{f}=\overline{\rho f}/\overline{\rho}$.
Fluctuations with respect to Reynolds and Favre averages are indicated by
$f'=f-\overline{f}$ and $f''=f-\widetilde{f}$,
respectively.

\section{Case description}
\label{sec:casedescription}
\begin{figure}
    \centering
    \includegraphics[width=\textwidth]{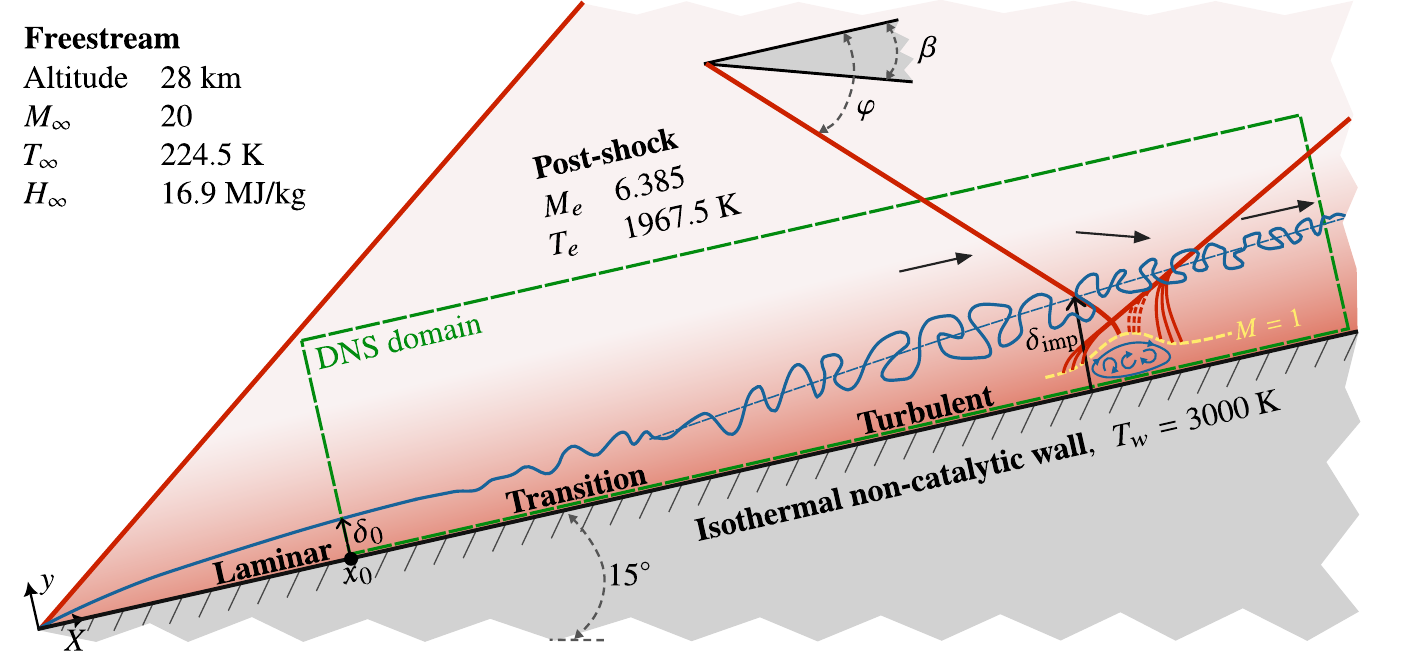}
    \caption{Sketch of the physical configuration under analysis.}
    \label{fig:illustration}
\end{figure}

The configuration considered in this work, illustrated in figure~\ref{fig:illustration}, consists of an oblique shock impinging on a turbulent boundary layer developing in the post-shock region of a wedge flying at hypersonic speed. Freestream conditions are representative of a suborbital flight at $M_\infty=20$ and an altitude of $28$ km for a $15^\circ$ wedge.
Post-shock boundary-layer edge conditions are summarised in table~\ref{tab:edgeconditions}. Under these conditions, air composition undergoes mild dissociation, leading to a non-negligible concentration of atomic and molecular species. The corresponding stagnation enthalpy is high enough for thermochemical effects to become prominent in regions where the flow is decelerated towards rest. In this process, viscous dissipation, i.e., the aerodynamic heating, is responsible for a consistent temperature rise, thereby activating high-enthalpy effects such as vibrational excitation and molecular dissociation.

\begin{table}
    \centering
    \begin{tabular}{c c c c c c c c c c c}
        $M_e$ & $\gamma_e$ & $T_e$ [K] & $\rho_e$ [kg/m$^3$] & $p_e$ [Pa] & $H_e$ [MJ/kg] & $Y_{\ce{N2}}$ & $Y_{\ce{O2}}$ & $Y_{\ce{NO}}$ & $Y_{\ce{O}}$ & $Y_{\ce{N}}$\\
        6.385 & 1.297 & 1967.5 & 0.135 & 76469.8 & 16.9 & 0.76363 & 0.22901 & 0.00721 & 0.00015 & 0.0 \\
    \end{tabular}
    \caption{Post-shock boundary-layer edge conditions.}
    \label{tab:edgeconditions}
\end{table}

In order to assess the role of thermochemical effects in hypersonic SBLI, three different fluid models are considered. The reference configuration is a fully reactive case (R), whose numerical treatment has been described in section~\ref{sec:methodology}. 
Two additional simplified models are introduced to progressively remove thermochemical effects while preserving the same external flow conditions.

In the thermally perfect case (TP), chemical reactions are suppressed, and the fluid is modelled as a single-species equivalent gas. Thermodynamic properties are temperature-dependent and are constructed to match those of the edge mixture of the reacting case.

In the calorically perfect gas case (CP), the same equivalent gas is considered, but with constant specific heats and a fixed ratio of specific heats equal to the freestream values of the previous configurations, $\gamma^{CP}=\gamma_e^{TP}=\gamma_e^{R}=1.297$. 

The comparison among the three cases is designed to isolate distinct physical effects.
Chemical non-equilibrium effects, including both changes in fluid properties and the activation of chemical reactions, are assessed by comparing the reactive (R) and thermally perfect (TP) cases.
Thermodynamic effects are instead quantified by comparing the thermally perfect (TP) and calorically perfect (CP) cases. This latter comparison is not intended to model vibrational non-equilibrium explicitly, since no separate vibrational temperature is solved. Rather, it isolates the sensitivity of the interaction to the caloric closure of the frozen gas model. The thermally perfect case accounts for the temperature dependence of the specific heats under a single-temperature equilibrium assumption, whereas the calorically perfect case keeps the specific heats, and hence $\gamma$, fixed at the edge value.
%This latter comparison is intended to bound the role of vibrational energy modes: the TP model assumes fully equilibrated vibrational degrees of freedom through temperature-dependent specific heats, whereas the CP model represents the opposite limit in which such modes are effectively frozen. %\textbf{can we say something like: The TP model accounts for temperature-dependent specific heats, thereby incorporating the equilibrium contribution of additional internal energy modes (notably vibrational modes) at elevated temperatures. In contrast, the CP model assumes constant specific heats, effectively neglecting such temperature-dependent contributions.}

All simulations share identical dimensional parameters, boundary conditions, and numerical setup. This choice is intended to mimic a controlled experimental configuration in which the external conditions are fixed, allowing the impact of the fluid model on both the incoming boundary layer and the SBLI to be directly quantified.

The computational domain corresponds to the green box shown in figure~\ref{fig:illustration}. The simulations are performed on a Cartesian grid covering a domain of size $L_x/\delta_0 \times L_y/\delta_0 \times L_z/\delta_0 = 1610 \times 64 \times 40$, where $\delta_0$ denotes the thickness of the laminar boundary layer at the inlet, defined as the wall-normal coordinate at which the streamwise velocity $u$ reaches 99\% of the edge velocity $u_e$.
The domain starts at a distance $x_0=56.9\delta_0$ from the leading edge. The streamwise coordinate used throughout the paper is measured from the inlet plane rather than from the leading edge, i.e., $x=X-x_0$, where $X$ denotes the distance from the leading edge. 

The lack of self-similarity in the species profiles prevents the use of turbulent inflow generators and therefore requires the simulation to cover the entire laminar-to-turbulent transition process. %\textbf{this is disputable, but ok} 
The laminar flow at the inflow is characterised by a Reynolds number based on the boundary layer thickness of $Re_{\delta_0}=\rho_e u_e \delta_0 / \mu_e=3800$. 
Laminar inflow profiles are generated by numerically solving the laminar boundary layer equation for a multispecies reacting mixture \citep{urzay2021direct,sciacovelli2021assessment}, with assumptions of frozen mixture and constant heat capacities for the TP and CP cases, respectively.
Transition is triggered via the application of a tripping volume force designed to mimic the effect of a roughness strip at the wall. This method is based on the approach of \cite{schlatter2012turbulent}, which consists of the addition of the following term in the momentum conservation equations:

\begin{equation}
    F_y(x,y,z,t) = \rho A_t \left\{\left[ 1-b\left(t\right)\right] h^i\left(z \right) + b\left(t \right) h^{i+1}\left(z \right) \right\} \exp \left[-\left(\frac{x-x_{tr}}{\ell_x} \right)^2-\left(\frac{y}{\ell_y} \right)^2\right]
\end{equation}

This represents a vertical volumetric force centred at $x_{tr}=20\delta_0$, $y_{tr}=0$, with length scales $\ell_x=4\delta_0$ and $\ell_y=\delta_0$ and amplitude $A_t=0.026 u_e^2/\delta_0$. This forcing is modulated both in time by $b(t)=3p^2-2p^3$ with $p(t)=t/t_s - \mathrm{floor}(t/t_s)$ and $t_s=4\delta_0/u_e$ and in the span by random harmonic signals $h^i(z)$ of unit amplitude and cut-off scale $z_s=1.7 \delta_0$. 

The impinging oblique shock is characterised by a deflection angle $\beta=8^\circ$ and a shock angle $\varphi=14.9^\circ$ and impinges at $x_\mathrm{imp}/\delta_0=1286$, well downstream of the transition region, where the boundary layer is fully turbulent (see \S~\ref{sec:blvalidation}).
From this point onward, the coordinates are expressed in terms of $\delta_\mathrm{imp}$ and $x_\mathrm{imp}$, so that $\hat{x}=(x-x_\mathrm{imp})/\delta_\mathrm{imp}$, $\hat{y}=y/\delta_\mathrm{imp}$, and $\hat{z}=z/\delta_\mathrm{imp}$.

At the top boundary and the outflow, non-reflecting boundary conditions are applied, and the oblique shock generator is simulated by enforcing inviscid shock relations at the top boundary to obtain the desired impinging location. Periodic boundary conditions are applied in the spanwise direction.

\begin{figure}
    \centering
    \includegraphics{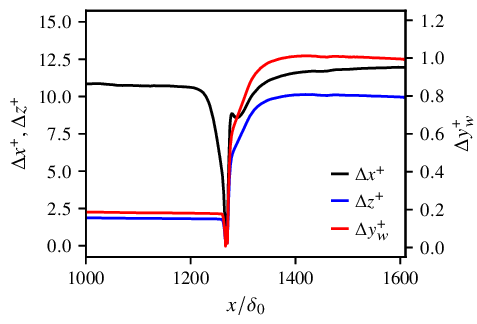}
    \caption{Streamwise evolution of the viscous-scaled grid spacings $\Delta x^+$, $\Delta z^+$, and $\Delta y_w^+$ for the reactive case.}
    \label{fig:viscspacings}
\end{figure}

The same grid is adopted for all cases, consisting of $15840 \times 520 \times 1280 \approx 10.5$ billion grid points. Stretching functions are applied in the streamwise direction to account for the reduction of the viscous length scale across the interaction region, and in the wall-normal direction to ensure a near-wall spacing $\Delta y^+_w \approx 1$ throughout the domain and a proper resolution in the entire boundary layer, while the grid is uniform in the spanwise direction. The streamwise evolution of the viscous-scaled spacings is shown in figure~\ref{fig:viscspacings} for the reactive case. Similar values are obtained for the remaining cases. The figure shows that the grid resolution remains within standard DNS requirements throughout the interaction region, where the viscous length scale decreases by approximately a factor of five.

Owing to the different thermodynamic models, the boundary layer reaches the impingement location with different states in the three cases. Considering the station $\hat{x}=-4$ as representative of the incoming boundary layer for the SBLI, the state of the boundary layer for the three configurations is summarised in table~\ref{tab:impblprop}.
\begin{table}
    \centering
    \begin{tabular}{l c c c c c c c c c}
        Case & $\delta/\delta_0$ & $Re_\tau$ & $Re_\theta$ & $Re_{\delta_2}$ & $Re_{\tau}^*$ & $\Theta$ & $M_\tau$ & $C_f \times 10^3$ & $-B_q$\\ 
        SBLI-R  & 16.2 & 972 & 3335 & 2604 & 1550 & 0.17 & 0.171 & 1.45 & 0.10 \\
        SBLI-TP & 15.4 & 983 & 3171 & 2590 & 1459 & 0.09 & 0.170 & 1.40 & 0.15\\
        SBLI-CP & 15.4 & 990 & 3161 & 2583 & 1470 & 0.09 & 0.168 & 1.40 & 0.15\\
    \end{tabular}
    \caption{Boundary layer averaged properties at $\hat{x}=-4$. Here, $Re_\tau=\rho_w u_\tau \delta/\mu_w$, $Re_\theta=\rho_e u_e \theta / \mu_e$, $Re_{\delta_2}=\rho_e u_e \theta / \mu_w$, $Re_{\tau}^*=\sqrt{\rho_e \tau_w} \delta / \mu_e$, $\Theta=(h_w-h_e)/(h_r-h_e)$, $M_\tau=u_\tau / \sqrt{\gamma_w R_w T_w}$, $C_f = \tau_w / (0.5 \rho_e u_e^2 )$, $B_q = q_w/(\rho_w u_\tau h_w)$ with $u_\tau=\sqrt{\tau_w/\rho_w}$, $h_r=h_e+ru_e^2/2$ and $r=0.9$, $\tau_w = (\mu \partial u/\partial y)|_w$ and $q_w = (\lambda \partial T / \partial y)|_w$}
    \label{tab:impblprop}
\end{table}

The wall is modelled as isothermal at $T_w=3000$ K and, for the reactive case, non-catalytic. Under the latter assumption, the wall does not participate in surface chemical reactions, i.e., $\partial Y_n / \partial y |_w = 0,\:n=1,...,N_S$. Although this assumption idealizes real thermal protection systems, which often rely on ablative materials such as resins, carbon composites, and silicone-based compounds, it considerably simplifies the numerical treatment and enables a cleaner comparison among the different simulations.
The adopted wall temperature is also higher than typical material limits (around $1500$ K), but is selected for two reasons. First, as shown by \citet{fratini2026wall}, chemical activity in hypersonic boundary layers increases with wall temperature; using a high $T_w$ therefore enhances thermochemical effects and makes their impact on key flow observables easier to identify. Second, simulations at a lower, more realistic $T_w$ would be substantially more expensive, especially for the present SBLI setup, where shock impingement markedly reduces the viscous length scale after the interaction and hence demands significantly finer resolution.
The present wall condition should therefore be interpreted as a controlled high-enthalpy numerical experiment rather than as a model of a specific thermal-protection material. Surface recombination and material-response effects are deliberately excluded, so that the gas-phase thermochemical response can be isolated more clearly. Consequently, the quantitative trends
reported below are conditional on the prescribed high-wall-temperature, non-catalytic setting.

The same wall temperature is prescribed in all cases, rather than matching non-dimensional parameters such as the diabatic parameter $\Theta=(h_w-h_e)/(h_r-h_e)$ or the Eckert number $Ec=u_e^2/(h_r-h_w)$, in order to mimic a controlled numerical experiment with fixed dimensional conditions. As a consequence, since dissociation products increase the formation enthalpy of the mixture, in the reactive case the flow reaches the wall at a higher enthalpy, thereby increasing the diabatic parameter $\Theta$. This implies that, at fixed wall temperature, the reactive case experiences weaker effective cooling, since part of the energy drain is already provided by endothermic chemical processes. Although several studies \citep{cogo2023assessment,wenzel2024heat,fratini2026wall} have highlighted the importance of wall-heat transfer in compressible wall-bounded flows, the values of $\Theta$ reported in table~\ref{tab:impblprop} remain relatively close to each other and, more importantly, fall within the same region of the wall-heat transfer regime diagram proposed by \citet{wenzel2024heat}. For this reason, differences associated with wall cooling in the incoming boundary layer are expected to play only a secondary role in the present comparison.

Each simulation was advanced for a total time of approximately $280 \delta_\mathrm{imp}/u_e$. Flow statistics and wall signals were collected only after the initial transient had decayed and the flow had reached a statistically steady state. The corresponding sampling window spans $160 \delta_\mathrm{imp}/u_e$. Spanwise-averaged statistics were sampled every $0.1 \delta_\mathrm{imp}/u_e$, whereas wall signals used for spectra analysis were stored with a finer sampling interval $0.01 \delta_\mathrm{imp}/u_e$. Based on the low-frequency timescale identified in the wall-pressure spectra, the available signal length corresponds to approximately 10 low-frequency cycles.

%Total simulation time $t u_e/\delta_{imp}$:
%circa 282
%
%Total time fot gahtering statistics and spectra:
%circa 163
%
%frequency for statistics (all cases)
%0.1
%
%frequency for spectra (all cases)
%0.01
%

\section{Turbulent boundary layer validation}
\label{sec:blvalidation}

Before introducing shock impingement, the incoming turbulent boundary layer is validated against reference data to ensure that the upstream state is accurately reproduced. Results from the reactive case only are shown for improved clarity, but validation has been carried out for the two remaining cases as well.

\begin{figure}
\centering
\begin{subfigure}{0.49\textwidth}
    \caption{}
    \includegraphics{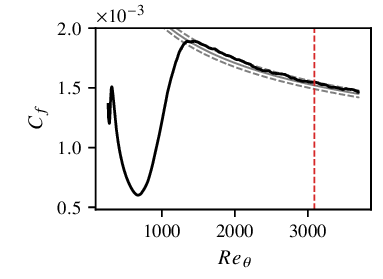}
    \label{fig:blvalidationcf}
\end{subfigure}
\hfill
\begin{subfigure}{0.49\textwidth}
    \caption{}
    \includegraphics{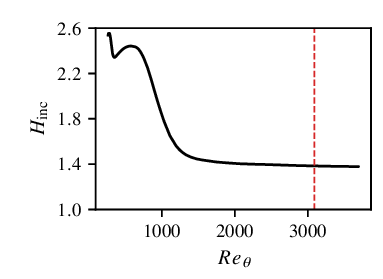}
    \label{fig:blvalidationincshapef}
\end{subfigure}
\caption{Streamwise evolution of the skin-friction coefficient (left panel) and of the incompressible shape factor (right panel) as a function of the momentum thickness Reynolds number. The solid gray line of panel (a) denotes the correlation $C_f = 0.0131 Re_\theta^{-0.268}$ from \cite{ceci2022numerical} and the dashed gray lines the $\pm 2 \%$ uncertainty band. The dashed red lines mark the station at which wall normal profiles are extracted. Data are shown only before the shock impingement.}
\label{fig:blvalidation1}
\end{figure}

First, the streamwise evolution of the skin-friction coefficient $C_f$ and of the incompressible shape factor $H_\text{inc}$ is shown in figure~\ref{fig:blvalidation1}. These are defined as

\begin{equation}
    C_f = \tau_w / (0.5 \rho_e u_e^2), \quad
    \tau_w = \overline{\mu \frac{\partial u }{\partial y }} \bigg|_w
\end{equation}

and

\begin{equation}
    H_\mathrm{inc} = \delta^*_\text{inc} / \theta_\text{inc}
\end{equation}

where the incompressible displacement and momentum thicknesses are

\begin{equation}
    \delta^*_\mathrm{inc} = \int_0^\delta \left( 1-\frac{u}{u_e}\right)dy, \quad \theta_\mathrm{inc} = \int_0^\delta \frac{u}{u_e}\left( 1-\frac{u}{u_e}\right)dy
\end{equation}

Inspection of figure~\ref{fig:blvalidation1} reveals the transition process from the laminar boundary layer imposed at the inflow to a fully developed turbulent state. The effects of the tripping volume force used to trigger transition are visible in the initial overshoot of $C_f$ and in a rapid drop of $H_\mathrm{inc}$. In the turbulent region, the skin-friction coefficient is in excellent agreement with the correlation proposed by \citet{ceci2022numerical}, lying well within the $\pm 2\%$ uncertainty band of the fit. The incompressible shape factor provides a complementary assessment: from the Blasius solution, a laminar boundary layer yields $H_\mathrm{inc} \approx 2.6$, while a 
canonical turbulent boundary layer settles around $H_\mathrm{inc} \approx 1.4$. The present data recover this asymptotic value in the turbulent region, confirming that the upstream boundary layer is in a fully turbulent state at the location selected for validation, indicated by the dashed line in figure~\ref{fig:blvalidation1}, located at $x/\delta_0=1016$. The corresponding flow properties at this reference station are reported in Table~\ref{tab:blprop_val}.
At the selected station, wall-normal mean velocity profiles and Reynolds-stresses are compared with canonical DNS datasets at comparable conditions. The reference cases are selected to match as closely as possible the semi-local friction Reynolds number $Re_\tau^*$, which accounts for the mean property variations across the boundary layer and represents the most appropriate parameter for comparison under compressibility and wall-cooling effects. Incompressible data from \citet{schlatter2012turbulent}, the high-Mach-number wall-cooled simulation of \citet{cogo2023assessment}, and the hypersonic turbulent boundary layer in chemical non-equilibrium of \citet{williams2025turbulence} are included, the latter also representing the closest available reference in terms of thermochemical state.
\begin{table}
    \centering
    \begin{tabular}{l c c c c c c c c c c c}
        Reference & $M_e$ & $Re_\tau$ & $Re_\theta$ & $Re_{\delta_2}$ & $Re_{\tau}^*$ & $\Theta$ & $M_\tau$ & $C_f \times 10^3$ & $-B_q$ & Reactive\\ 
        Present work & 6.4 & 843 & 3087 & 2312 & 1426 & 0.18 & 0.18 & 1.55 & 0.10 & YES\\
        \citet{schlatter2012turbulent} & - & 1272 & 4061 & 4061 & 1272 &  - & - & 2.97 & - & NO\\
        %\citet{pirozzoli2011turbulence} & 2 & 843 & 4431 & 2825 & 1682 & 1 & 0.07  & 2.28 & 0 & NO\\
        \citet{cogo2023assessment} & 6 & 444 & 2675 & 1313 & 1466 & 0.25 & 0.16 & 1.40 & 0.09 & NO\\
        \citet{williams2025turbulence} & 7 & 1175 & 3155 & 2836 & 1483 & 0.15 & 0.19 & 1.60 & 0.12 & YES\\
    \end{tabular}
    \caption{Boundary layer averaged properties at the station selected for validation.}
    \label{tab:blprop_val}
\end{table}

\begin{figure}
\centering
\begin{subfigure}{0.49\textwidth}
    \caption{}
    \includegraphics{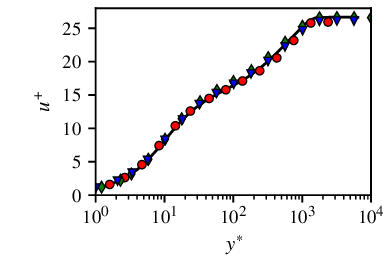}
    \label{fig:blvalidationmeanvelprof}
\end{subfigure}
\hfill
\begin{subfigure}{0.49\textwidth}
    \caption{}
    \includegraphics{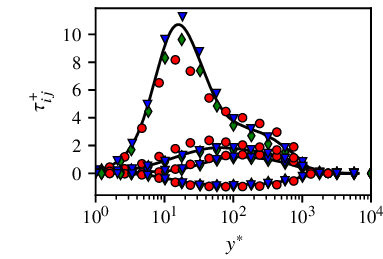}
    \label{fig:blvalidationreystress}
\end{subfigure}
\caption{Validation of the turbulent boundary-layer statistics against literature data. Left panel: Hasan-transformed mean velocity profile. Right panel: density-scaled Reynolds-stress components. The solid line denotes the present reactive simulation; symbols denote reference data from \citet{schlatter2012turbulent} (red circles), \citet{cogo2023assessment} (green diamonds), and \citet{williams2025turbulence} (blue triangles).}
\label{fig:blvalidation2}
\end{figure}

Figure~\ref{fig:blvalidation2} reports the Hasan-transformed mean velocity profile (left panel) and the density-scaled Reynolds-stress components (right panel). Both quantities are plotted against the wall-normal coordinate in semi-local units $y^*=y/\delta_v^*$ (introduced by \citet{huang1995compressible}), with $\delta_v^*=\mu/\sqrt{\rho \tau_w}$, which has proven a better collapse of both mean and fluctuating velocity profiles from different flow conditions, especially when strong variations in the flow properties are present in the boundary layer. The mean velocity profile shows very good agreement from the viscous sublayer to the logarithmic region, indicating that the near-wall momentum balance is properly captured.

The density-scaled Reynolds-stress, $\tau_{ij}^+ = (\bar{\rho}/\bar{\rho}_w) \widetilde{u_i'' u_j''} / u_\tau^2$, distributions are also consistent with literature trends. As expected, the largest deviations are observed with the incompressible dataset of \citet{schlatter2012turbulent}. A closer agreement is instead obtained with the Mach 6 cooled data from \citet{cogo2023assessment} and with the reactive simulation of \citet{williams2025turbulence}, which better reproduces the present thermodynamic conditions.
Overall, the validation confirms that the inflow boundary layer is statistically converged and suitable for assessing the effects of fluid-thermochemical modelling in the subsequent SBLI analysis. 

\section{Reactive simulation}
\label{sec:reactivesimulation}
The interaction between the incoming turbulent hypersonic boundary layer and the impinging oblique shock results in a relatively weak shock/boundary layer interaction characterised by incipient flow separation.
Despite the high Mach number, the combined effect of moderate Reynolds number and strong wall cooling limits the extent of the separated region, in agreement with previous observations in high-speed SBLI configurations \citep{volpiani2018effects,volpiani2020effects}. The separation and reattachment locations are found at $\hat{x}=-1.226$ and $\hat{x}=-0.992$, respectively, resulting in a relatively small recirculation bubble.
\begin{figure}
    \centering
    \includegraphics{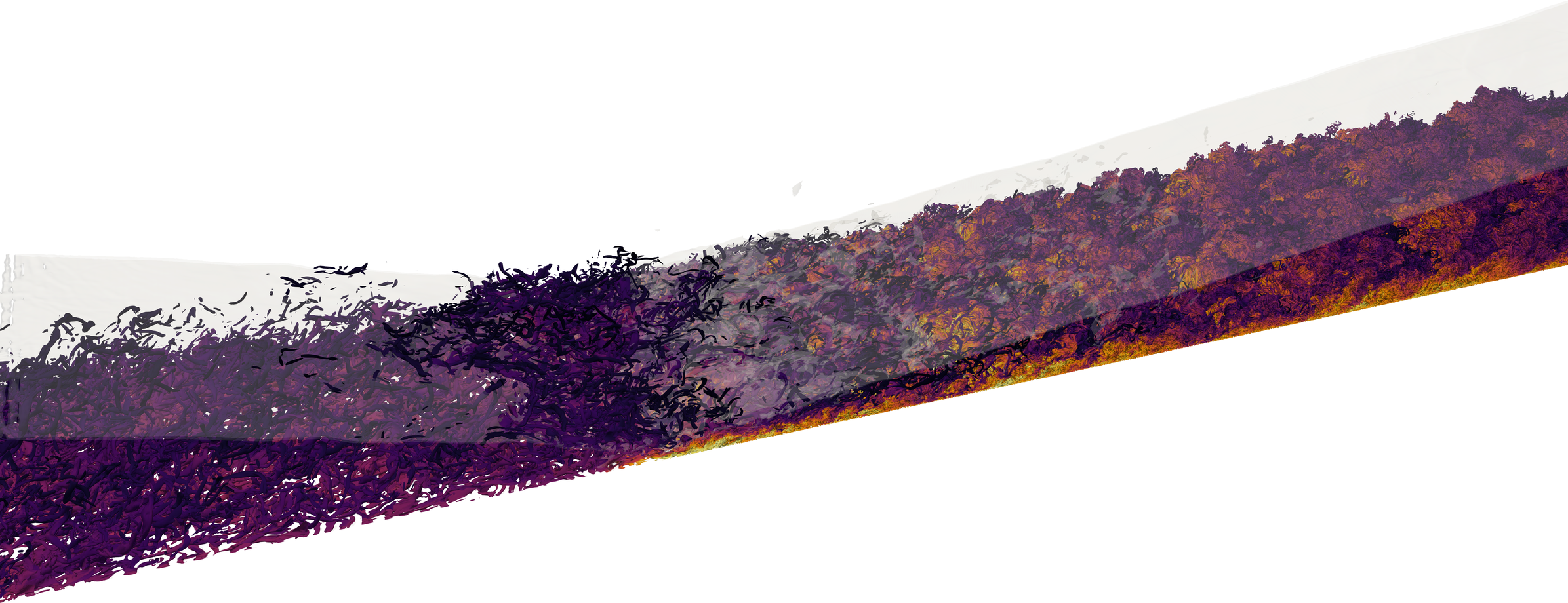}
    \includegraphics{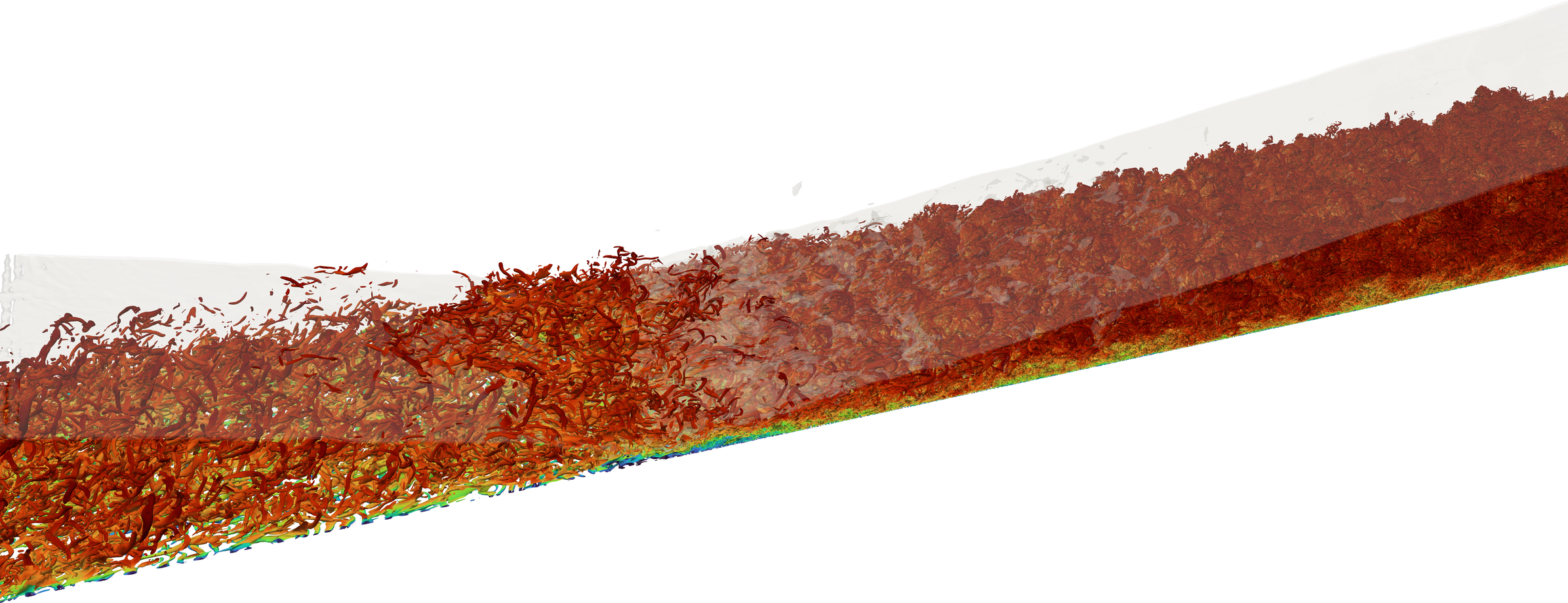}
    \caption{Instantaneous three-dimensional visualisation of the simulation. Light gray: isosurface of the shock sensor revealing the position of both the impinging and the reflected shocks. Isosurfaces of the swirling strength are coloured by (top panel) the atomic oxygen mass fraction with range $0 \leq Y_{\ce{O}} \leq 0.15$, and (bottom panel) streamwise velocity with range $0 \leq u/u_e \leq 1$.}
    \label{fig:snapshot}
\end{figure}
\begin{figure}
    \centering
    \begin{subfigure}{0.49\textwidth}
        \caption{}
        \includegraphics{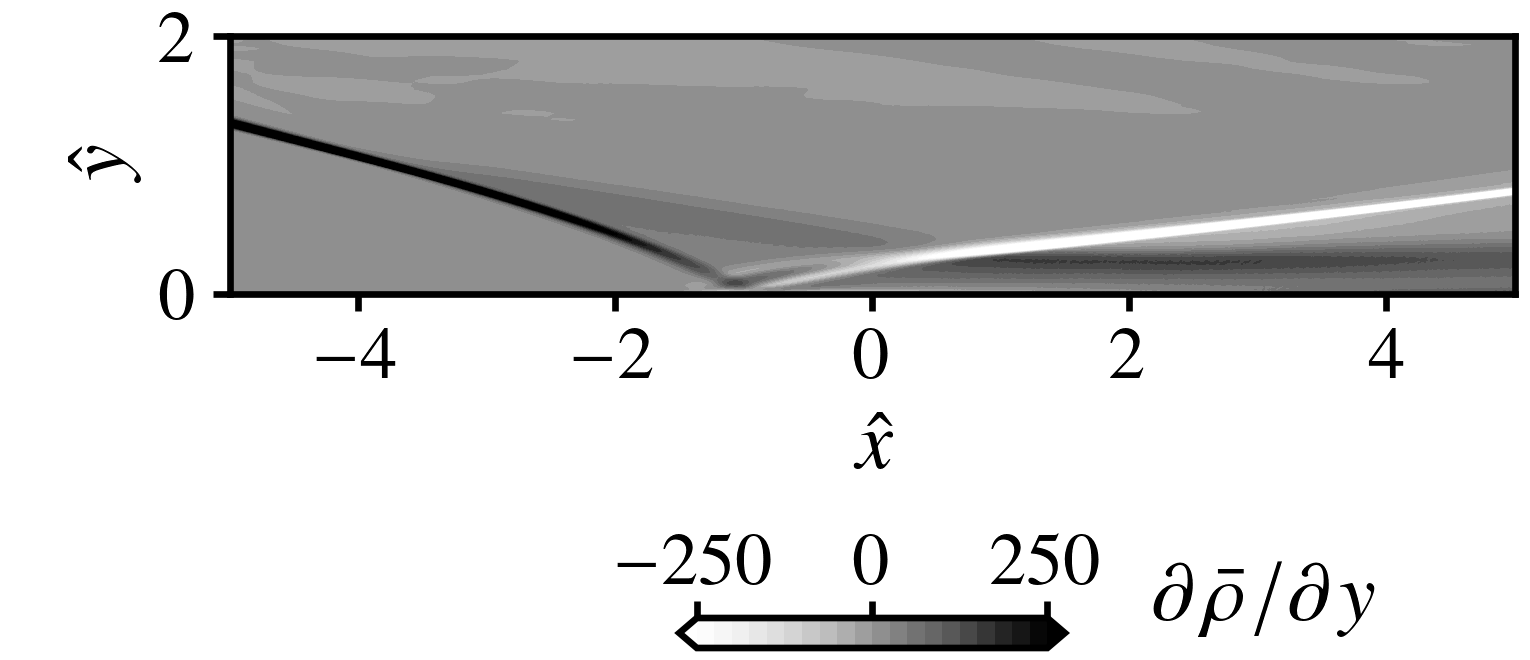}
        \label{fig:mean_rhograd}
    \end{subfigure}
    \begin{subfigure}{0.49\textwidth}
        \caption{}
        \includegraphics{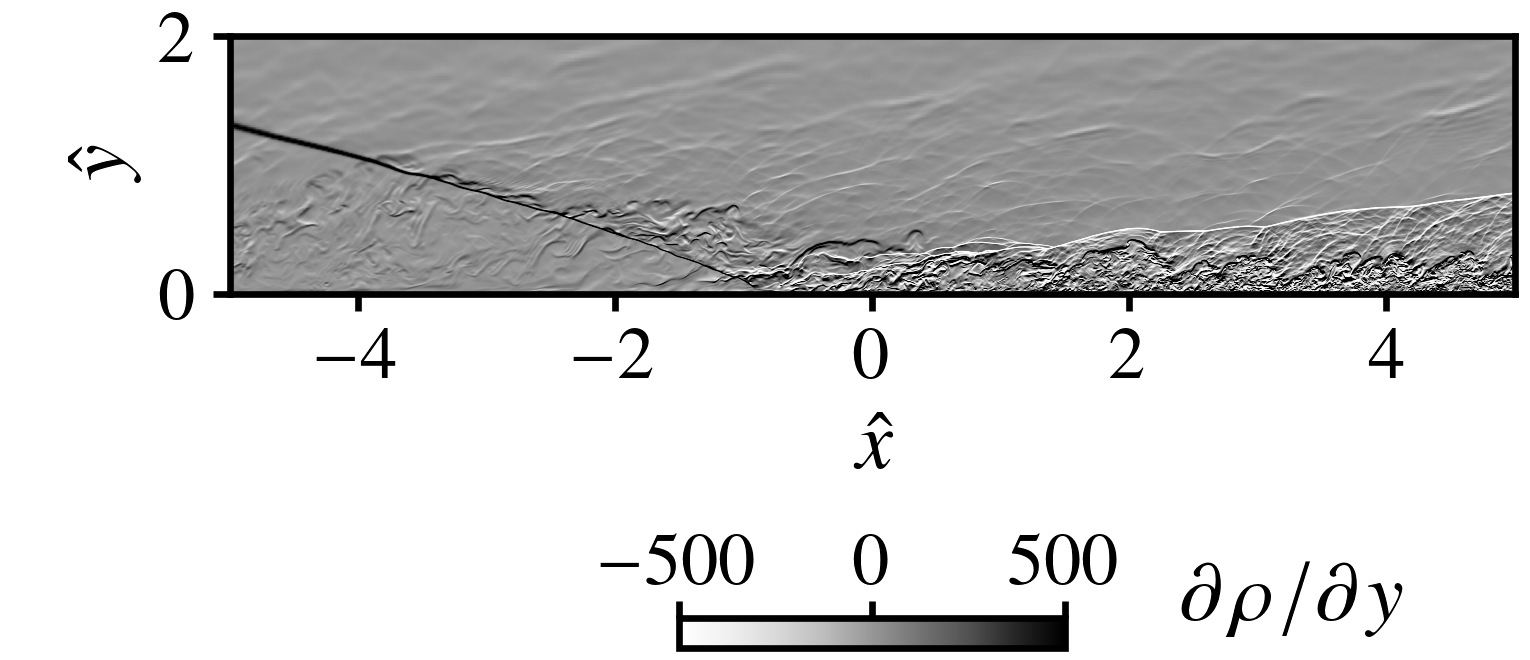}
        \label{fig:inst_rhograd}
    \end{subfigure}
    \hfill
    \begin{subfigure}{0.49\textwidth}
        \caption{}
        \includegraphics{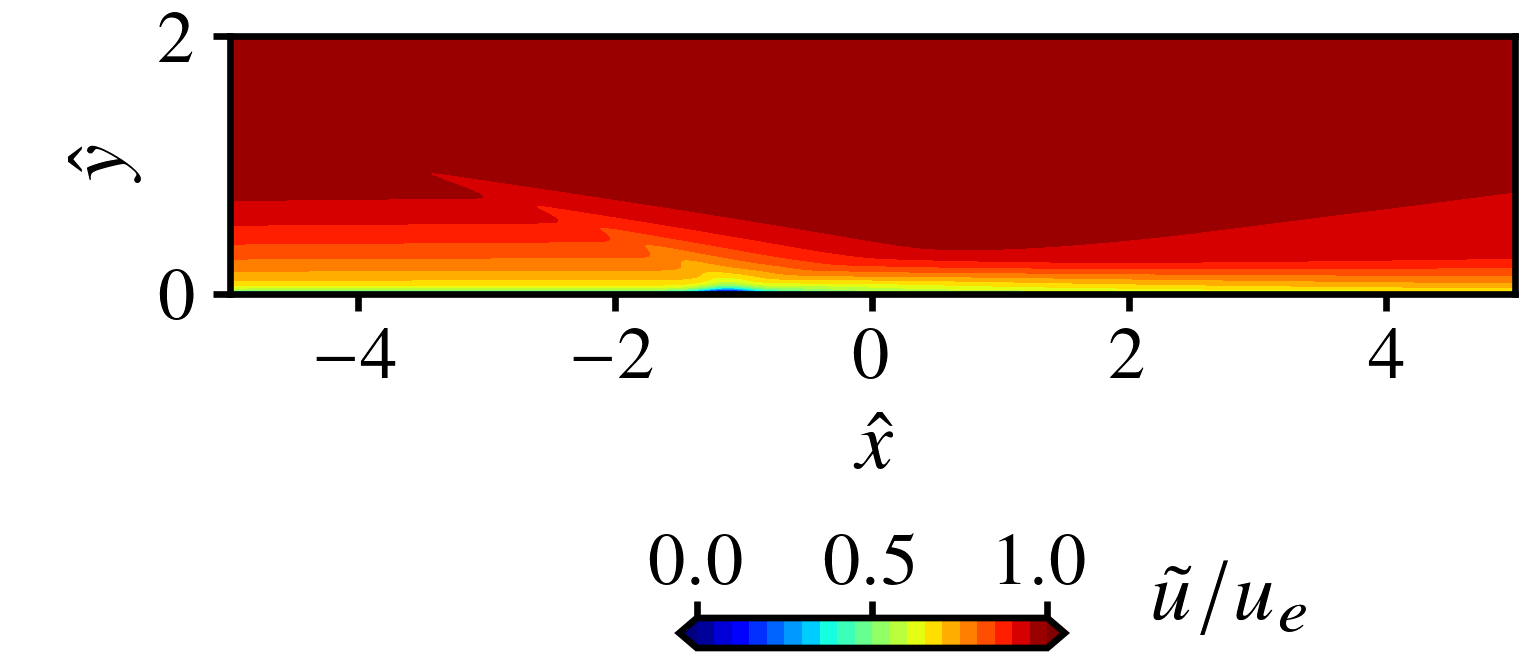}
        \label{fig:mean_u}
    \end{subfigure}
    \begin{subfigure}{0.49\textwidth}
        \caption{}
        \includegraphics{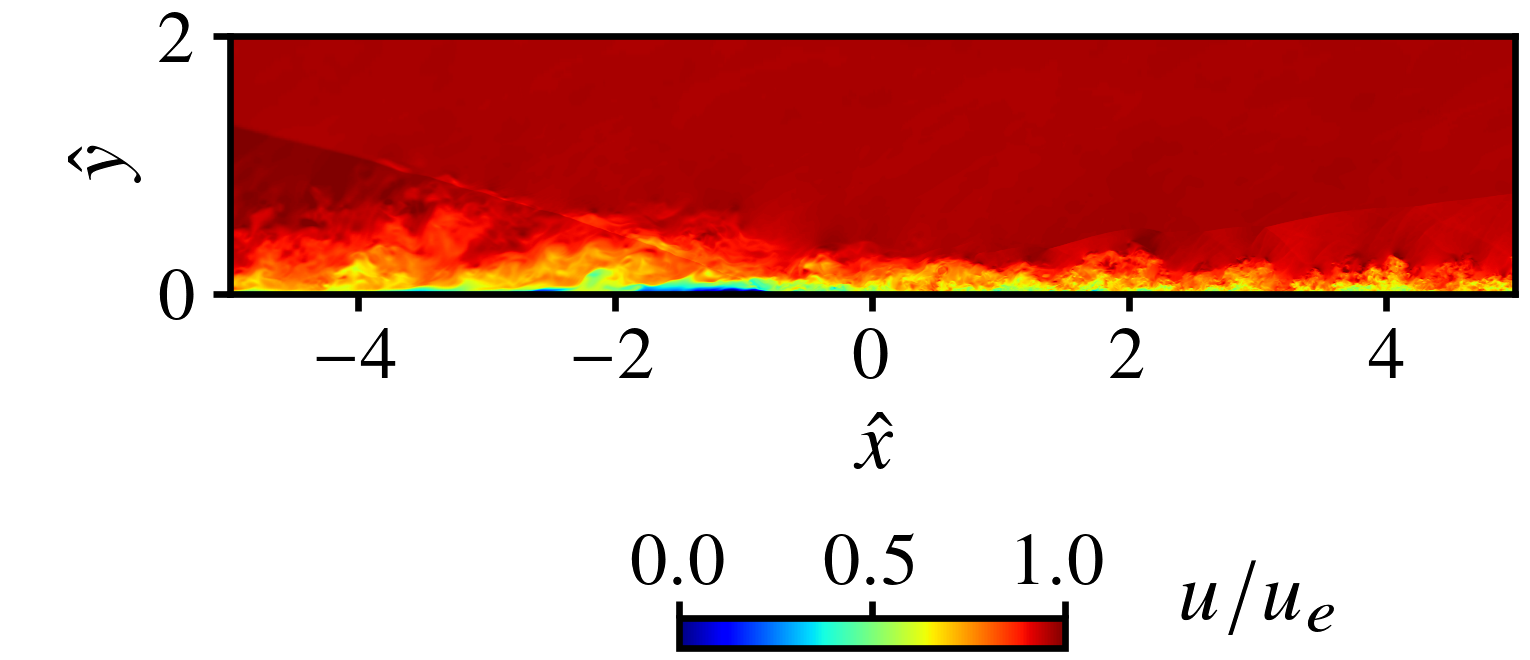}
        \label{fig:inst_u}
    \end{subfigure}
    \hfill 
    \begin{subfigure}{0.49\textwidth}
        \caption{}
        \includegraphics{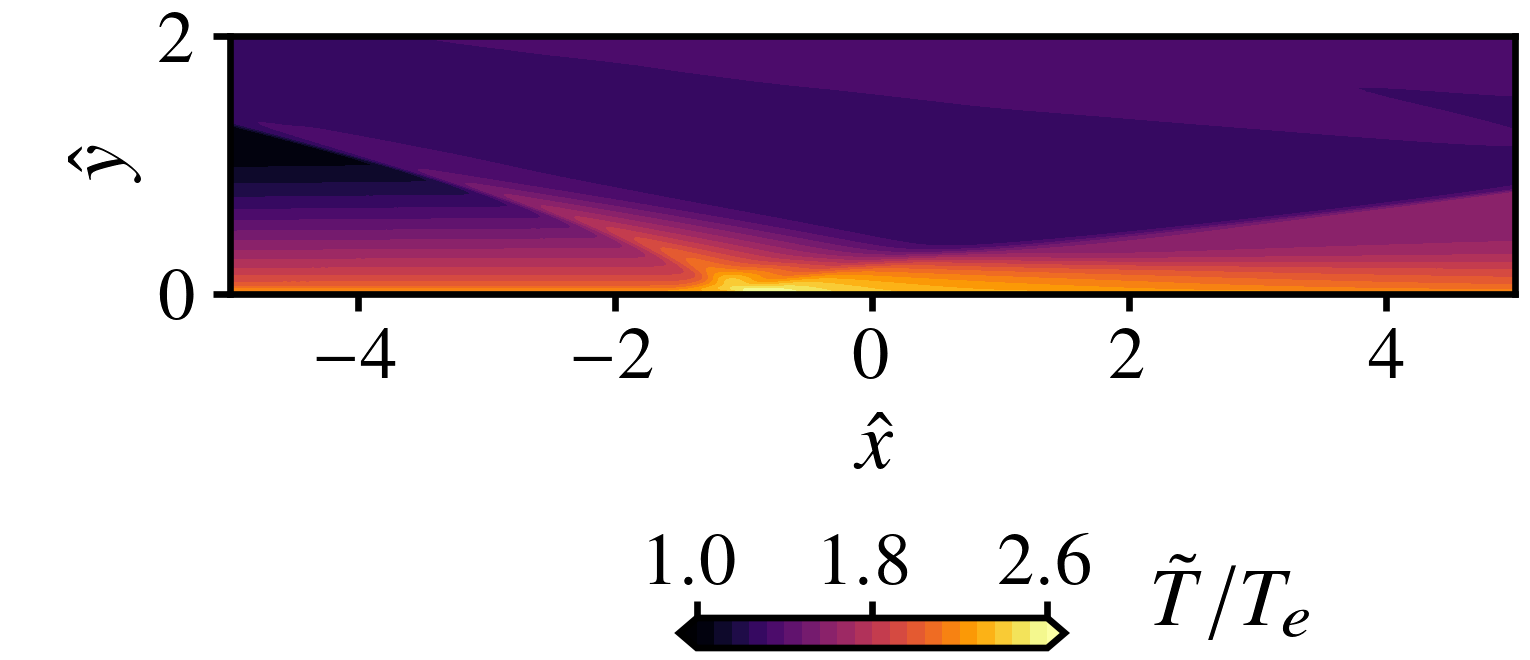}
        \label{fig:mean_t}
    \end{subfigure}
    \begin{subfigure}{0.49\textwidth}
        \caption{}
        \includegraphics{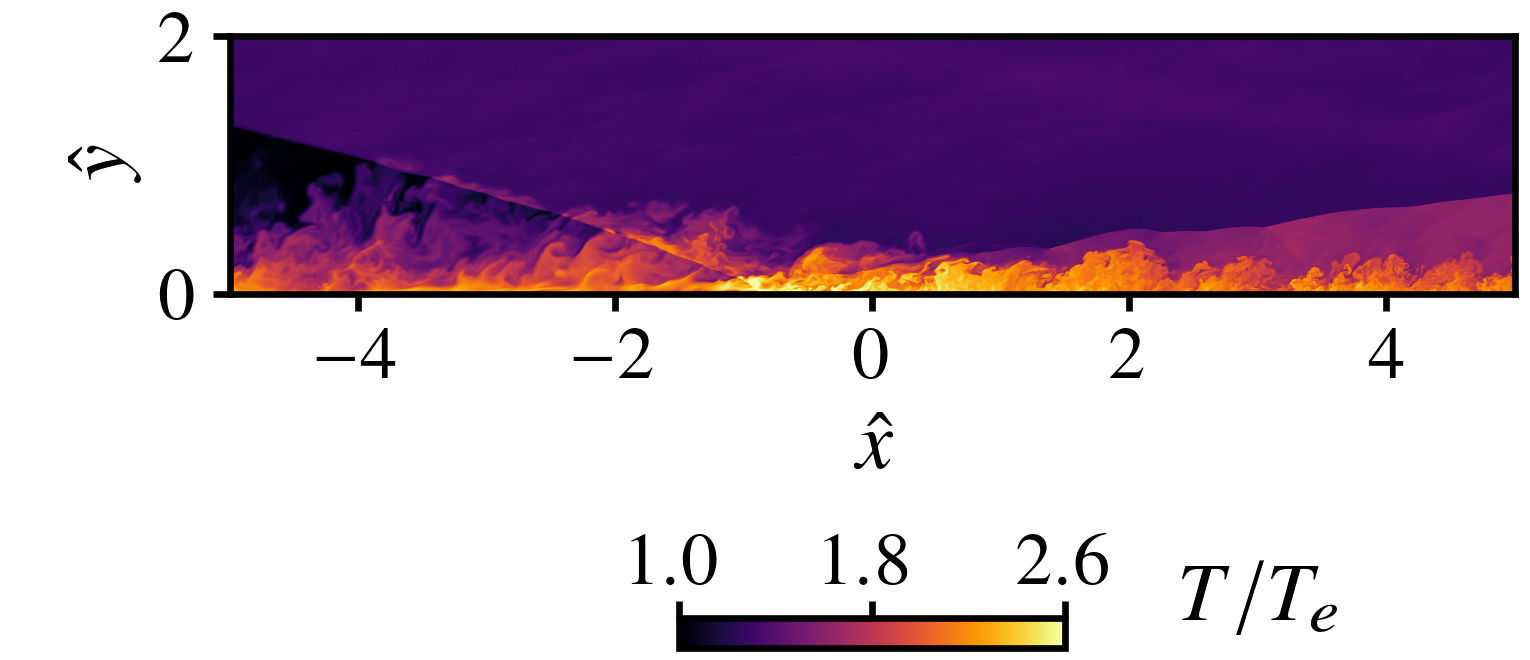}
        \label{fig:inst_t}
    \end{subfigure}
    \hfill
    \begin{subfigure}{0.49\textwidth}
        \caption{}
        \includegraphics{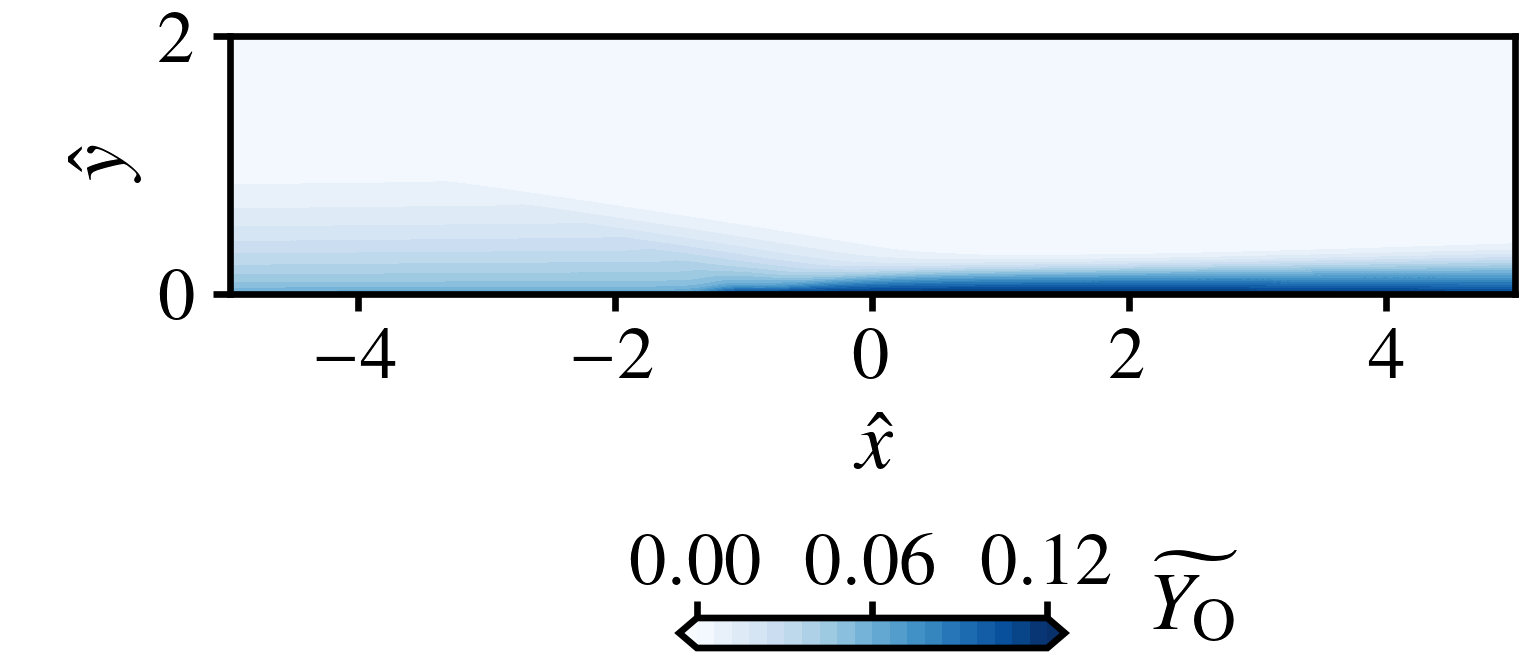}
        \label{fig:mean_y0}
    \end{subfigure}
    \begin{subfigure}{0.49\textwidth}
        \caption{}
        \includegraphics{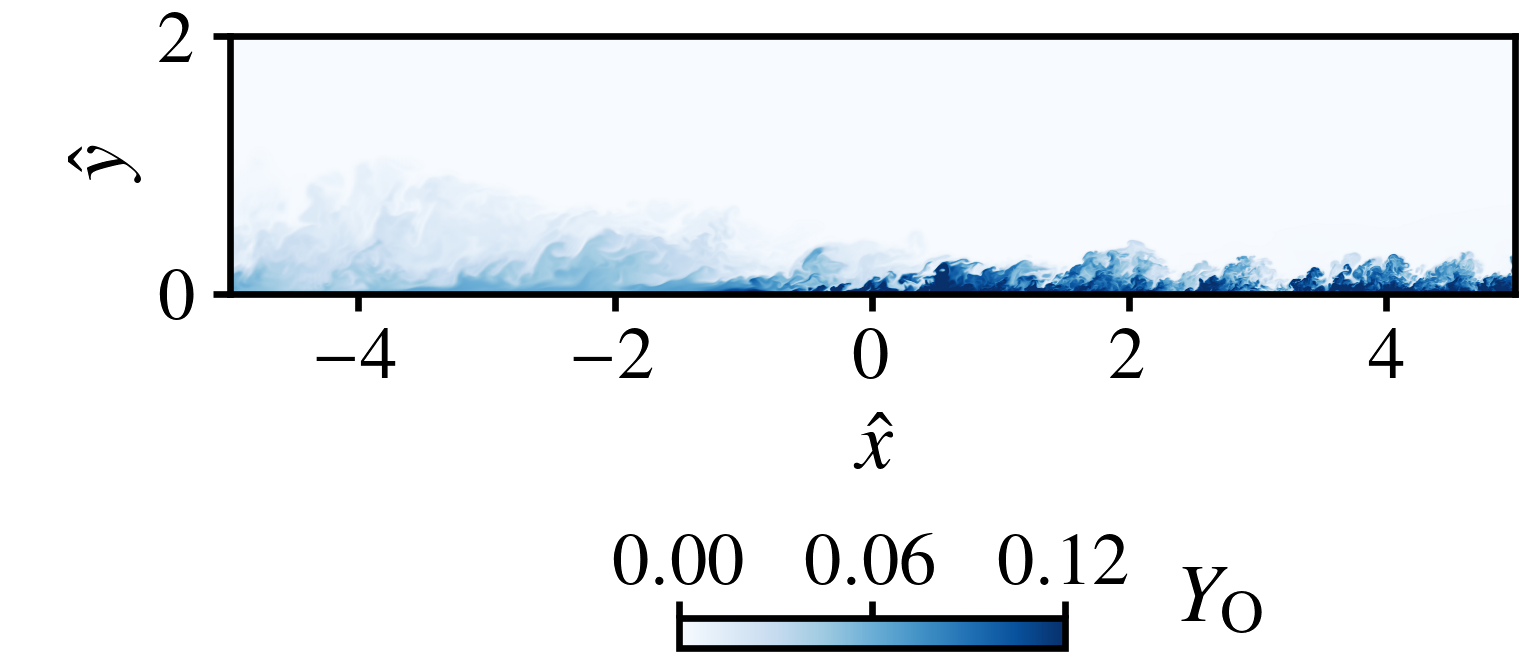}
        \label{fig:inst_y0}
    \end{subfigure}
    \caption{Mean and instantaneous $x-y$ slices of the reactive simulation. (a-b) Density gradient in the $y$ direction, (c-d) streamwise velocity, (e-f) temperature, and (g-h) atomic oxygen mass fraction.}
    \label{fig:xyslices}
\end{figure}

The overall topology of the interaction is qualitatively illustrated in figures~\ref{fig:snapshot} and~\ref{fig:xyslices}. The three-dimensional instantaneous visualisation (figure~\ref{fig:snapshot}) indicates that turbulent structures become smaller and more densely packed across the interaction, reflecting the strong compression induced by the shock system. 
This compression is accompanied by a marked temperature increase and, consequently, by an increase in the concentration of the radical $Y_{\ce{O}}$.
The combined instantaneous and mean $x$--$y$ slices of figure~\ref{fig:xyslices} provide a clearer view of the shock pattern and of the associated thermochemical response. The mean flow organisation is dominated by the impinging and reflected shocks, whereas the separation shock remains weak and rapidly merges with the reflected one, as visible in the mean density-gradient field (figures~\ref{fig:xyslices}(a-b)). The corresponding velocity field, shown in figures~\ref{fig:xyslices}(c-d), confirms this picture. The mean streamwise velocity is deflected across the impinging and reflected shocks, with the high-momentum region progressively compressed against the wall downstream of the interaction. The instantaneous field highlights the intensified small-scale turbulent activity that characterises the post-interaction boundary layer.
This compression of the boundary layer downstream of the interaction is a direct consequence of the relatively weak slope of both the impinging and reflected shocks, which is a distinctive feature of the present hypersonic configuration. In particular, the impinging shock has an angle $\varphi=14.9^\circ$, whereas the reflected shock is inclined by $8.27^\circ$. 
As a result, the boundary layer is confined between the wall and the reflected shock, so that the post-interaction boundary layer is thinner than the incoming one. This behaviour differs from the thickening more commonly observed in lower Mach number SBLI. In addition, the weak slope of the reflected shock causes it to interact strongly with the outer turbulent structures, producing visible corrugations, consistently with the observations of \citet{direnzo2024stagnation}.

The thermal response is also clearly visible in figures~\ref{fig:xyslices}(e-f). The interaction induces a substantial temperature rise throughout the boundary layer, with peak values reaching approximately $T \approx 4500$ K in the interaction region. The increase in temperature is accompanied by a marked enhancement of thermochemical activity, as shown by the corresponding increase in the atomic-oxygen mass fraction, visible in figures~\ref{fig:xyslices}(g-h). Figure~\ref{fig:xyslices}, therefore, already suggests that the SBLI does not simply compress the boundary layer, but also drives it into a regime of significantly stronger chemical activity.

\begin{figure}
    \centering
    \begin{subfigure}{0.49\textwidth}
        \caption{}
        \includegraphics{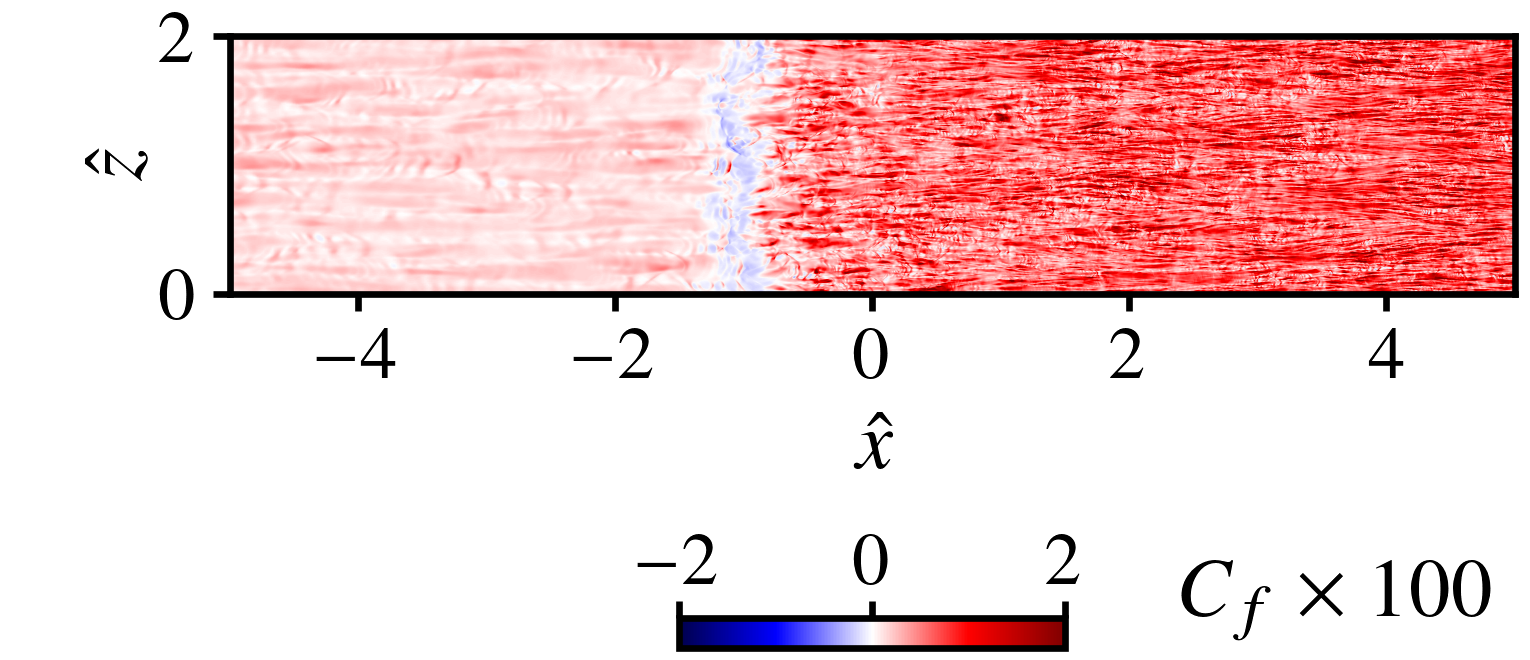}
        \label{fig:inst_cf}
    \end{subfigure}
    \begin{subfigure}{0.49\textwidth}
        \caption{}
        \includegraphics{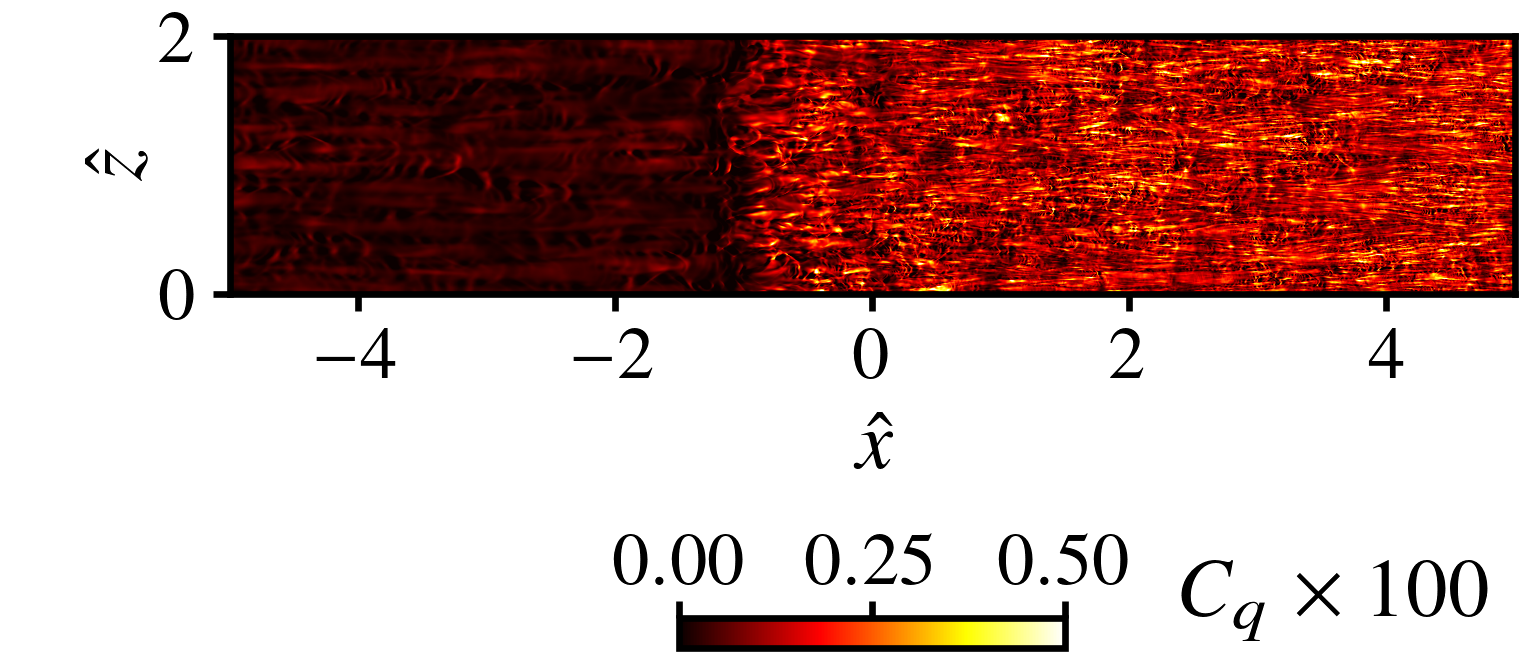}
        \label{fig:inst_cq}
    \end{subfigure}
    \caption{Instantaneous slices at the wall of the reactive simulation. (a) skin friction coefficient, and (b) wall heat flux coefficient.}
    \label{fig:wallinst}
\end{figure}

The instantaneous wall features of the interaction are shown in
figure~\ref{fig:wallinst}. The skin-friction field clearly identifies the separation region through the appearance of negative values, while both wall shear and wall heat flux exhibit a marked spanwise modulation associated with the footprint of the turbulent structures interacting with the shock system. Downstream of the interaction, both quantities increase significantly, in agreement with the compressed and thermally activated state of the post-interaction boundary layer.

To gain more detailed insight into the thermal and chemical response to the SBLI, three representative streamwise stations are selected, and the corresponding wall-normal profiles are analysed. The chosen stations are $\hat{x}=-4$, upstream of the SBLI and representative of the incoming boundary layer; $\hat{x}=-1.1$, approximately at the centre of the separation bubble; $\hat{x}=4$, downstream of the interaction. 

\begin{figure}
    \centering
    \begin{subfigure}{0.49\textwidth}
        \caption{}
        \includegraphics{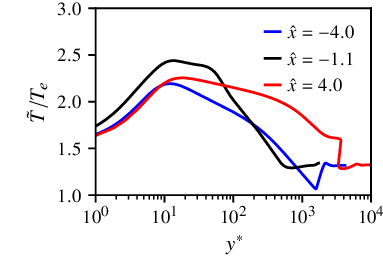}
        \label{fig:meantemprofiles}
    \end{subfigure}
    \begin{subfigure}{0.49\textwidth}
        \caption{}
        \includegraphics{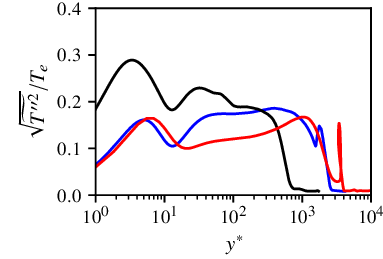}
        \label{fig:trmsprofiles}
    \end{subfigure}
    \caption{Wall-normal profiles of (a) Favre-averaged temperature and (b) temperature fluctuations.}
    \label{fig:temprofiles}
\end{figure}

The wall-normal temperature profiles, shown in figure~\ref{fig:temprofiles}(a), provide a first quantitative view
of the thermal response. Upstream of the interaction, the mean temperature profile exhibits the classical structure of a moderately cooled compressible boundary layer, with a temperature peak located in the buffer region around $y^*\approx15$, consistently with previous DNS studies \citep{cogo2023assessment,wenzel2024heat,williams2025turbulence,fratini2026wall}.
Within the interaction region, the temperature increases across the entire boundary layer, whereas further downstream the profile becomes thicker and slightly cooler as a result of thermal diffusion and molecular dissociation. Temperature fluctuations, shown in figure~\ref{fig:temprofiles}(b), display the canonical double-peaked structure of cooled high-speed boundary layers and are strongly amplified within the interaction region, reaching values of up to about 30\% of the freestream temperature, reflecting the increase in turbulence activity induced by the shock. Downstream of the interaction, fluctuation levels decrease, consistent with the reduced gradients associated with the broadened thermal profile.

The thermal field alone does not provide a complete picture of the interaction, as finite-rate chemistry introduces a delayed response with respect to the rapid temperature variations induced by the shock.  

\begin{figure}
    \centering
    \begin{subfigure}{0.49\textwidth}
        \caption{}
        \includegraphics{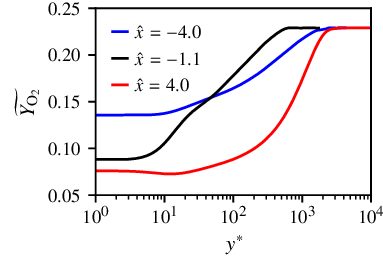}
        \label{fig:o2profiles}
    \end{subfigure}
    \begin{subfigure}{0.49\textwidth}
        \caption{}
        \includegraphics{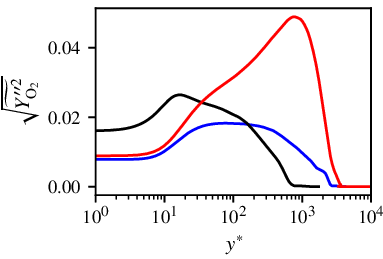}
        \label{fig:o2rmsprofiles}
    \end{subfigure}
    \hfill
    \begin{subfigure}{0.49\textwidth}
        \caption{}
        \includegraphics{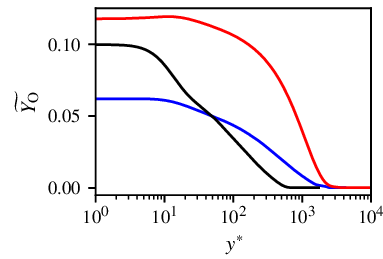}
        \label{fig:oprofiles}
    \end{subfigure}
    \begin{subfigure}{0.49\textwidth}
        \caption{}
        \includegraphics{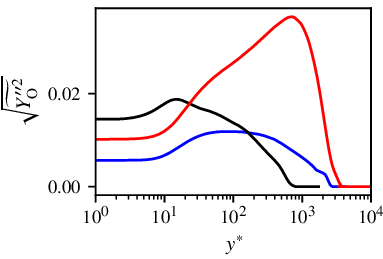}
        \label{fig:ormsprofiles}
    \end{subfigure}
    \hfill
    \begin{subfigure}{0.49\textwidth}
        \caption{}
        \includegraphics{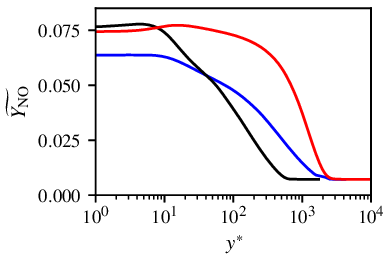}
        \label{fig:noprofiles}
    \end{subfigure}
    \begin{subfigure}{0.49\textwidth}
        \caption{}
        \includegraphics{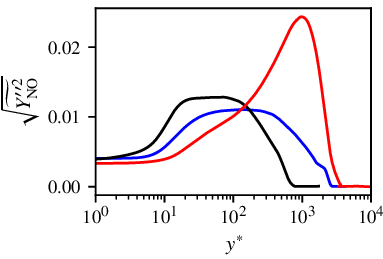}
        \label{fig:normsprofiles}
    \end{subfigure}
    \caption{Wall-normal profiles of the Favre-averaged (left column) and root-mean-square (right column) species mass fractions at $\hat{x}=-4$, $-1.1$, and $4$ for (a,b) \ce{O2}, (c,d) \ce{O}, and (e,f) \ce{NO}. The wall-normal coordinate is expressed in semi-local units.}
    \label{fig:speciesprofiles}
\end{figure}
\begin{figure}
    \centering
    \includegraphics{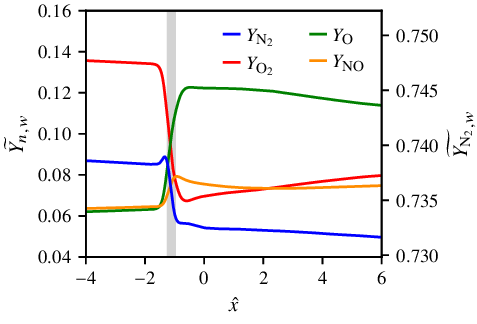}
    \caption{Streamwise distribution of species mass fraction at the wall for \ce{N2}, \ce{O2}, \ce{O}, and \ce{NO}. The light gray shaded region indicates the separation bubble. }
    \label{fig:wallcomposition}
\end{figure}

The wall-normal distributions of Favre-averaged species mass fractions and their root-mean-square fluctuations are shown in figure~\eqref{fig:speciesprofiles} for the (a) \ce{O2}, (c) \ce{O}, and (e) \ce{NO}, and provide a clearer assessment of chemical non-equilibrium. 
The production of atomic oxygen is primarily driven by oxygen dissociation (step \eqref{eq:R2}), whereas nitric oxide is formed through the forward shuffle reaction \eqref{eq:R4} and the backward shuffle reaction \eqref{eq:R5} (Zel'Dovich mechanism).
Molecular nitrogen is not shown due to its weak participation in chemical kinetics, the same applies to atomic nitrogen, which is mildly produced near the temperature peak from forward reactions \eqref{eq:R1} and \eqref{eq:R4} and is quickly depleted by the backward step \eqref{eq:R5}.
Upstream of the interaction, the species profiles exhibit the canonical structure of hypersonic turbulent boundary layers  \citep{urzay2021direct,williams2025turbulence,fratini2026wall}, with dissociation products (in particular atomic oxygen and nitric oxide) increasing from the logarithmic region towards the wall, plummeting near the temperature peak, and then diffusing towards the wall. Within the interaction region, although temperature rises sharply, the corresponding increase is comparatively weak. In particular, the maximum in the temperature profiles observed in
figure~\ref{fig:temprofiles}(a) does not correspond to a maximum in chemical
activity. Instead, peak concentrations of radicals within the boundary layer are attained further downstream of the interaction.
This mismatch between thermal and chemical response to the SBLI clearly indicates that chemical processes are not able to instantaneously adjust to the rapid thermal forcing imposed by the shock. The streamwise evolution of the wall mass fractions shown in figure~\ref{fig:wallcomposition} confirms this interpretation.
The wall species mass fractions exhibit a delayed response with respect to the interaction region. Chemical activity starts near the separation point but develops over a streamwise extent that exceeds the length of the interaction. In particular,  atomic oxygen reaches its peak wall concentration well after the reattachment point, while nitric oxide attains its maximum close to reattachment, further supporting the presence of strong chemical non-equilibrium.

Additional insight is provided by the root-mean-square (r.m.s.) of species mass fraction fluctuations, shown in figures~\ref{fig:speciesprofiles} (b), (d), and (f).
In the incoming boundary layer, turbulent mixing between the external cold, undissociated air and the hotter, dissociated air from the inner region leads to peak fluctuation levels in the logarithmic region. Within the interaction region, r.m.s. levels are significantly amplified, particularly in the near-wall region, as a result of the strong thermodynamic fluctuations induced by the SBLI.
Further downstream, the fluctuation levels reach their maximum at the boundary layer edge. This behaviour reflects the effects of the enhanced chemical activity, which steepens species gradients in the outer layer, promoting turbulent mixing and therefore increasing fluctuations.

\begin{figure}
    \centering
    \begin{subfigure}{0.49\textwidth}
        \caption{}
        \includegraphics{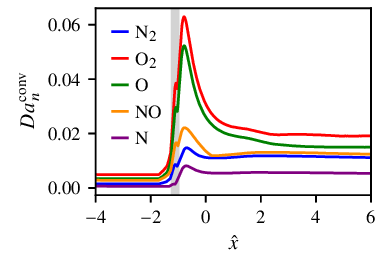}
        \label{fig:DaConv}
    \end{subfigure}
    \begin{subfigure}{0.49\textwidth}
        \caption{}
        \includegraphics{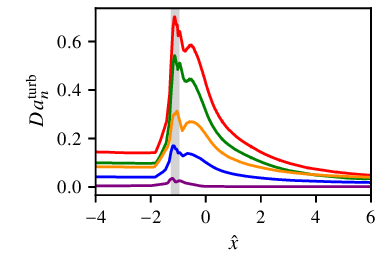}
        \label{fig:DaT}
    \end{subfigure}
    \caption{Streamwise distribution of the (a) convective and (b) turbulent Damk\"{o}hler numbers for all species. The light gray shaded region indicates the separation bubble.}
    \label{fig:Damkohlers}
\end{figure}

To quantify the degree of non-equilibrium, convective and turbulent Damk\"{o}hler numbers can be evaluated for each species $n$ as
$Da^{\mathrm{conv}}_n = \max_y \left(\tau^{\mathrm{conv}}/\tau^{\mathrm{chem}}_n \right)$
and 
$Da^{\mathrm{turb}}_n = \max_y \left(\tau^{\mathrm{turb}}/\tau^{\mathrm{chem}}_n \right)$, 
where $\tau^{\mathrm{conv}}=\delta_\mathrm{imp}/u_e$ is the convective time scale, $\tau^{\mathrm{turb}}=\kappa/\varepsilon$ is the turbulent time scale, and $\tau^{\mathrm{chem}}_n=\rho/|\dot{\omega}_n|$ is the chemical time scale of species $n$, with $\kappa=\widetilde{u_i''u_i''}/2$ and $\varepsilon=-\overline{\sigma_{ij}' \partial u_i'' / \partial x_j}$. The convective Damk\"{o}hler compares chemical time scales with an outer convective time scale based on the incoming boundary layer thickness at impingement, while the turbulent Damk\"{o}hler number quantifies the relative importance of turbulence–chemistry interaction. Their streamwise evolution is shown in figure~\ref{fig:Damkohlers}.
Consistent with other DNS studies of hypersonic turbulent boundary layers, upstream of the interaction the convective Damk\"{o}hler numbers remain well below unity, indicating that chemical kinetics are significantly slower than the characteristic time-scale of the boundary layer. The SBLI induces a marked increase in $Da^{\mathrm{conv}}_n$, with peak values approximately one order of magnitude larger than in the incoming boundary layer, particularly for atomic and molecular oxygen. Notably, the maximum values are attained downstream of the interaction region, further supporting the interpretation of a lagged chemical response to the rapid temperature rise induced by the shock. The turbulent 
Damk\"{o}hler number exhibits a different behaviour, being already moderately high in the upstream boundary layer and increasing up to $\mathcal{O}(1)$ within the interaction region. This indicates that chemistry operates on time scales closer to those of turbulence, especially within the recirculation region, where thermodynamic fluctuations are strongest. As a result, turbulence–chemistry interaction becomes increasingly relevant within the interaction region.

\begin{figure}
    \centering
    \includegraphics{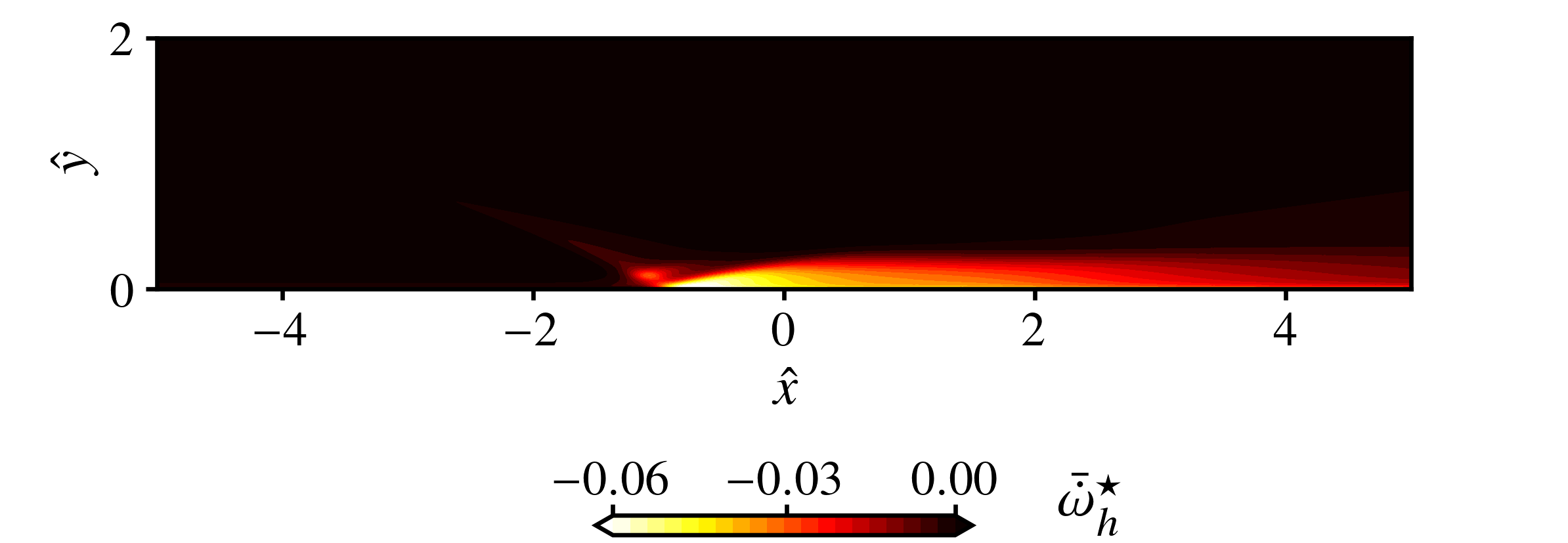}
    \caption{Mean $x-y$ slice of the normalised heat release rate 
$\bar{\dot{\omega}}_h^\star = \bar{\dot{\omega}}_h \, \delta_\mathrm{imp} / (\rho_e u_e H_e)$, where $\bar{\dot{\omega}}_h=-\sum_{n=1}^{N_s} \dot{\omega}_n \Delta h_{f,n}^{T_{\mathrm{ref}}}$.}
    \label{fig:hrr}
\end{figure}

The energetic footprint of the finite-rate chemistry is further illustrated in figure~\ref{fig:hrr}, which reports the mean chemical heat-release rate

\begin{equation}
\bar{\dot{\omega}}_h^\star =
\bar{\dot{\omega}}_h \delta_{\rm imp}/(\rho_e u_e H_e),
\qquad
\bar{\dot{\omega}}_h
=
-\overline{\sum_{n=1}^{N_s}
\dot{\omega}_n \Delta h^{T_{\rm ref}}_{f,n}} .
\end{equation}

With this convention, negative values denote endothermic chemical activity, namely the conversion of sensible energy into chemical formation enthalpy. The field shows that the SBLI strongly amplifies this endothermic contribution, which remains mainly concentrated in the interaction region where the shock-induced temperature rise promotes molecular dissociation. Non-negligible negative values persist downstream of the separation bubble, consistently with the delayed increase of dissociation products shown in figures~\ref{fig:speciesprofiles} and~\ref{fig:wallcomposition}. This supports the interpretation that the chemical response is not only compositional, but also energetic: finite-rate chemistry provides an additional pathway through which part of the shock-induced thermal energy is stored in chemical form.

\begin{figure}
\centering
\includegraphics{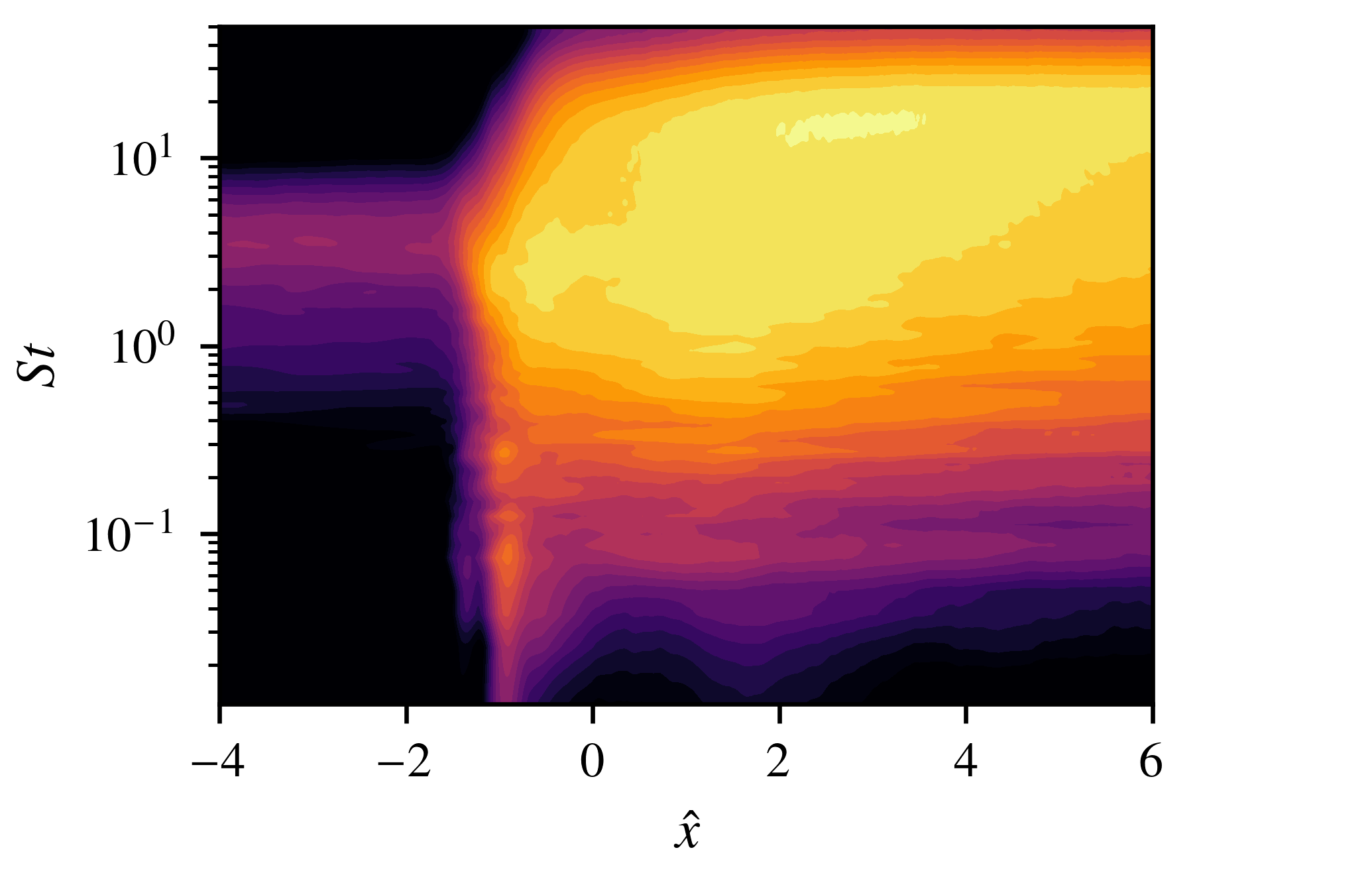}
\caption{Streamwise distribution of the spanwise-averaged premultiplied wall-pressure spectrum, $f\:\Phi_{pp}(f)$. The Strouhal number is defined as $St=f \delta_\mathrm{imp}/u_e$.}
\label{fig:pwspectrum}
\end{figure}

A characteristic feature of SBLI is the presence of low-frequency unsteadiness. Although its physical origin is not yet fully clarified, it is commonly associated with a breathing motion of the separation bubble and the accompanying drift of the shock system. The intensity and spectral content of this low-frequency dynamics depend on the interaction strength and on the flow conditions \citep{bernardini2023unsteadiness,quadros2018numerical}, and provide an additional time scale against which the relevance of finite-rate chemistry can be assessed. Figure~\ref{fig:pwspectrum} reports the streamwise distribution of the spanwise-averaged premultiplied wall-pressure spectrum. Upstream of the interaction, the spectral energy is concentrated at high frequencies, consistent with the canonical turbulent boundary-layer dynamics. Across the interaction, a distinct broadband low-frequency content emerges, with a peak located at a Strouhal number  $St^{\mathrm{LF}} \approx 0.07$. Downstream of the interaction, the energy content spans mid to high frequencies and is noticeably amplified relative to the upstream level, consistent with enhanced turbulent activity. 
The time-scale associated with the low-frequency dynamics, $\tau^{\mathrm{LF}}=\delta_\mathrm{imp}/\left(u_e St^{\mathrm{LF}}\right) \approx 14 \tau^{\mathrm{conv}}$, provides a further reference for chemical activity. Defining a low-frequency Damk\"{o}hler number as $Da_n^{\mathrm{LF}}=\max_y\left(\tau^{\mathrm{LF}}/\tau^{\mathrm{chem}}_n\right) \approx 14 Da_n^{\mathrm{conv}}$, values of order unity are recovered for the most chemically active species. This suggests that, although chemistry is relatively slow compared with the convective time-scale of the flow, its time scales are comparable to those of the low-frequency shock unsteadiness, raising the possibility of a non-trivial coupling between the two. A detailed characterisation of this coupling, however, falls outside the scope of the present work. The sampled signal covers approximately ten low-frequency cycles, which is sufficient to identify the dominant time-scale, but not to characterise the feedback mechanism in detail. A dedicated investigation of the chemistry-unsteadiness interplay is therefore left for future work.

%\begin{table}
%\centering
%\label{tab:damkohler}
%\begin{tabular}{r cc cc cc cc cc}
% & \multicolumn{2}{c}{\ce{N2}} & \multicolumn{2}{c}{\ce{O2}} & \multicolumn{2}{c}{\ce{O}} & \multicolumn{2}{c}{\ce{NO}} & \multicolumn{2}{c}{\ce{N}} \\
%$\hat{x}$ & $Da_c$ & $Da_t$ & $Da_c$ & $Da_t$ & $Da_c$ & $Da_t$ & $Da_c$ & $Da_t$ & $Da_c$ & $Da_t$ \\
%$-4.0$ &  0.117 & 0.042 & 0.389 & 0.145 & 0.270 & 0.100 & 0.223 & 0.084 & 0.050 & 0.004 \\
%$-1.1$ &  0.704 & 0.157 & 3.089 & 0.686 & 2.383 & 0.530 & 1.330 & 0.303 & 0.128 & 0.021 \\
%$4.0$  &  1.004 & 0.023 & 1.692 & 0.073 & 1.310 & 0.049 & 1.115 & 0.050 & 0.483 & 0.001 \\
%\end{tabular}
%\caption{Convective and turbulent Damk\"{o}hler numbers at selected streamwise stations.}
%\end{table}

\section{Effects of fluid modelling}
\label{sec:fluidmodeling}
This section assesses the impact of fluid-model simplifications on the SBLI by comparing the reactive (SBLI-R), thermally perfect (SBLI-TP), and calorically perfect (SBLI-CP) simulations. The objective is to separate the effects of finite-rate chemistry from those associated with the thermal model. It should be recalled that, although the external flow conditions are kept fixed, the incoming boundary layer reaches the impingement location with slightly different states in the three cases, as summarised in table~\ref{tab:impblprop}. At the same time, the incoming boundary layers remain close to each other, especially in terms of mean velocity profile, so that only a limited effect of the upstream state on the subsequent interaction would be expected.

\begin{figure}
\centering
\begin{subfigure}{0.49\textwidth}
    \caption{}
    \includegraphics{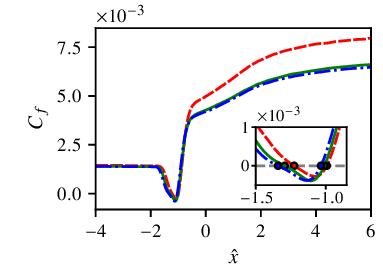}
    \label{fig:Cf}
\end{subfigure}
\hfill
\begin{subfigure}{0.49\textwidth}
    \caption{}
    \includegraphics{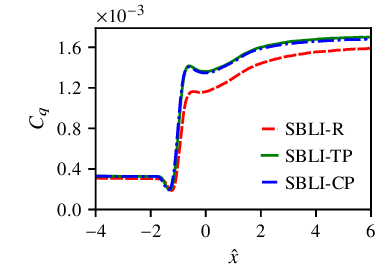}
    \label{fig:Cq}
\end{subfigure}
\begin{subfigure}{0.49\textwidth}
    \caption{}
    \includegraphics{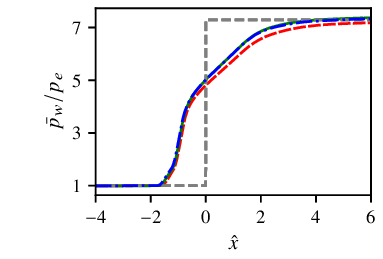}
    \label{fig:pw}
\end{subfigure}
\hfill
\begin{subfigure}{0.49\textwidth}
    \caption{}
    \includegraphics{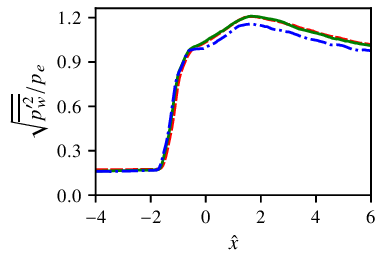}
    \label{fig:prms}
\end{subfigure}
\caption{Streamwise distribution of wall quantities for the reactive (SBLI-R), thermally perfect (SBLI-TP), and calorically perfect (CP) simulations. (a) skin-friction coefficient, (b) wall heat flux coefficient, (c) wall pressure scaled with the edge value, and (d) root-mean-square of wall pressure fluctuations. The inset in panel (a) provides a magnified view of the interaction region. The dashed gray line in panel (c) indicates the theoretical inviscid pressure jump.}
\label{fig:cfcqpwpwrms}
\end{figure}

The streamwise distributions of the skin friction coefficient $C_f$, wall heat flux coefficient $C_q=q_w/(\rho_e u_e^3)$, wall pressure, and r.m.s of wall pressure fluctuations are reported in figure~\ref{fig:cfcqpwpwrms}, while the main interaction properties are listed in table~\ref{tab:SBLI_params}. 

\begin{table}
    \centering
    \begin{tabular}{l c c c c c}
        Case & $\hat{x}_\mathrm{sep}$ & $\hat{x}_\mathrm{reatt}$ & $L_\mathrm{sep}/\delta_\mathrm{imp}$ & $\varphi_\mathrm{imp}$ & $\varphi_\mathrm{refl}$\\
        SBLI-R  & $-1.226$ & $-0.992$ & $0.233$ & $14.9^\circ$ & $8.27^\circ$ \\
        SBLI-TP & $-1.293$ & $-1.009$ & $0.284$ & $14.9^\circ$ & $8.49^\circ$ \\
        SBLI-CP & $-1.341$ & $-1.034$ & $0.307$ & $14.9^\circ$ & $8.49^\circ$ \\
    \end{tabular}
    \caption{Separation and shock parameters of the reactive (SBLI-R), thermally perfect (SBLI-TP), and calorically perfect (SBLI-CP) interactions.}
    \label{tab:SBLI_params}
\end{table}

\begin{figure}
\centering
\begin{subfigure}{0.49\textwidth}
    \caption{}
    \includegraphics{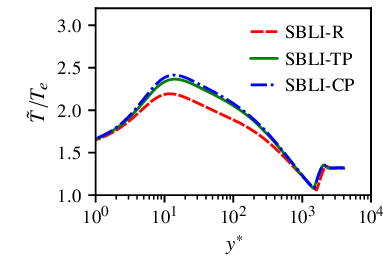}
    \label{fig:t_t0_0}
\end{subfigure}
\hfill
\begin{subfigure}{0.49\textwidth}
    \caption{}
    \includegraphics{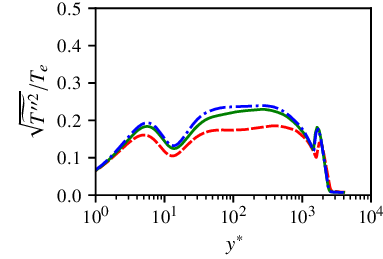}
    \label{fig:trms_t0_0}
\end{subfigure}
\begin{subfigure}{0.49\textwidth}
    \caption{}
    \includegraphics{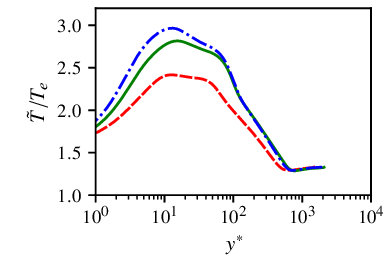}
    \label{fig:t_t0_1}
\end{subfigure}
\hfill
\begin{subfigure}{0.49\textwidth}
    \caption{}
    \includegraphics{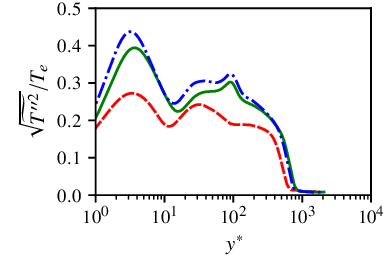}
    \label{fig:trms_t0_1}
\end{subfigure}
\begin{subfigure}{0.49\textwidth}
    \caption{}
    \includegraphics{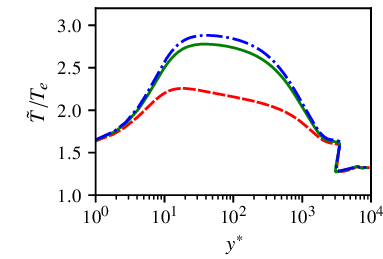}
    \label{fig:t_t0_2}
\end{subfigure}
\hfill
\begin{subfigure}{0.49\textwidth}
    \caption{}
    \includegraphics{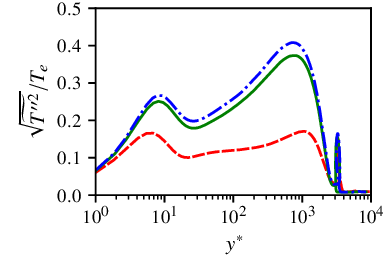}
    \label{fig:trms_t0_2}
\end{subfigure}
\caption{Wall-normal profiles of Favre-averaged temperature (left column) and
root-mean-square temperature fluctuations (right column) for the reactive
(SBLI-R), thermally perfect (SBLI-TP), and calorically perfect (SBLI-CP)
simulations.
Panels (a,b), (c,d), and (e,f) correspond to the stations $\hat{x}=-4$,
$-1.1$, and $4$, respectively.}
\label{fig:t_trms_stations}
\end{figure}

A first general observation is that the thermally perfect and calorically perfect simulations yield very similar predictions for all the considered wall observables. The difference between these two cases is limited to a slightly delayed separation onset in SBLI-CP and minor variations in the downstream thermal field. This close agreement indicates that, within the class of single-temperature frozen-gas closures considered here, the present SBLI is only weakly sensitive to the details of the caloric closure. This result should not be interpreted as a direct assessment of vibrational non-equilibrium, since no separate vibrational temperature is solved. By contrast, the reactive simulation exhibits large deviations from both frozen cases, indicating that finite-rate chemistry has a non-negligible effect on the overall interaction.
A first consequence of chemical non-equilibrium is the reduction of the separation bubble length. As shown in table~\ref{tab:SBLI_params}, the bubble extent decreases from $L_{\mathrm{sep}}/\delta_{\mathrm{imp}}=0.307$ in SBLI-CP and $0.284$ in SBLI-TP to $0.233$ in SBLI-R, corresponding to a reduction of
approximately $18\%$ relative to SBLI-TP. This difference is mainly associated with a downstream shift of the separation location, whereas the reattachment point is only weakly affected. 
Part of this trend can likely be attributed to differences in the incoming boundary-layer state, although the incoming velocity profiles remain overall very similar in the three cases. At the same time, the reduced bubble size observed in the reactive case is consistent with the known stabilising effect of wall cooling on SBLI \citep{volpiani2018effects,volpiani2020effects}. In the present reactive case, dissociation introduces an additional endothermic energy sink within the interaction region, so that part of the shock-induced thermal energy is absorbed by chemical processes. The resulting reduction of the near-wall thermal load is expected to weaken the interaction and contribute to the smaller separation bubble. 
While the absolute differences remain moderate, this conclusion should be interpreted in the light of the relatively weak interaction considered here, since stronger interactions may exhibit a more pronounced sensitivity to the fluid model.
Downstream of the reattachment, SBLI-R also shows larger skin friction values than the frozen cases. This difference is likely associated with the combined effect of composition-dependent transport properties and the modified thermal field, both of which affect the local Reynolds number and therefore the wall shear stress. Indeed, as shown by the mean wall-normal temperature profiles reported in figure~\ref{fig:t_trms_stations}, the reactive simulation displays systematically lower mean temperature than the other cases, especially within and downstream of the interaction region. This behaviour is consistent with the activation of endothermic dissociation reactions. By contrast, the thermally perfect and calorically perfect simulations remain relatively close to each other at all stations, with SBLI-CP generally reaching slightly higher temperatures owing to its lower thermal inertia. 

The wall heat-flux distributions shown in figure~\ref{fig:cfcqpwpwrms}(b) are fully consistent with this thermal behaviour. SBLI-TP and SBLI-CP provide nearly identical predictions, whereas SBLI-R yields systematically lower values downstream of the interaction. Since the wall heat flux is directly related to the wall-normal temperature gradient, the lower thermal level of the reactive boundary layer naturally results in a weaker heat-transfer response. In the present configuration, simplified frozen models therefore tend to over-predict the downstream thermal loads. Note that the wall heat transfer is expressed through the coefficient $C_q$ rather than through the Stanton number $C_h=q_w/\left[\rho_e u_e (h_r-h_w)\right]$. In the reactive case, the wall enthalpy is not constant along the streamwise direction, owing to changes in the composition. The use of $C_h$ would therefore introduce a streamwise dependence in the normalisation, making the comparison among the different cases less direct.

The effect of chemistry is also visible in the wall-pressure distribution, reported in figure~\ref{fig:cfcqpwpwrms}(c). All cases exhibit the typical signature of a weak interaction, with a first rise associated with the separation shock and a second, stronger compression through the reflected shock, without the extended plateau characteristic of stronger SBLI. More interestingly, SBLI-TP and SBLI-CP recover the theoretical inviscid pressure jump downstream of the interaction, whereas SBLI-R remains below it. The increased chemical activity and the associated variation in the composition change the local ratio of specific heats, which in turn weakens the reflected shock and slightly reduces its angle (see table~\ref{tab:SBLI_params}). The lower wall pressure in the reactive case is therefore a direct consequence of the thermodynamic
modification induced by finite-rate chemistry. 

Wall pressure fluctuations, shown in figure~\ref{fig:cfcqpwpwrms}(d), display a different pattern. In this case, SBLI-R and SBLI-TP remain very close over most of the interaction and downstream region, whereas SBLI-CP predicts slightly lower fluctuation levels.

Wall-normal profiles of temperature fluctuations are shown in figures~\ref{fig:t_trms_stations}(b), (d), and (f). At all three streamwise stations, the fluctuation levels follow the same hierarchy already observed for the mean temperature field, with SBLI-CP exhibiting the largest values, SBLI-TP intermediate ones, and SBLI-R the lowest. This ordering is consistent with the different thermal inertia of the three models. The calorically perfect case, having constant thermal inertia, develops the largest fluctuations, whereas in the thermally perfect case, part of the energy is lost by the increase of the specific heat capacities with temperature. The reactive case shows the highest damping, with the additional thermal sink associated with chemical reactions. 

\begin{figure}
\centering
\begin{subfigure}{0.49\textwidth}
    \caption{}
    \includegraphics{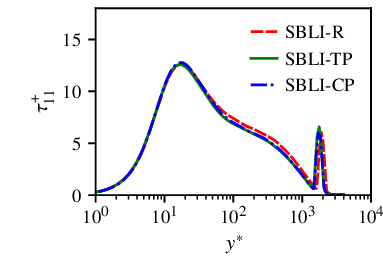}
    \label{fig:resystressx_st0}
\end{subfigure}
\hfill
\begin{subfigure}{0.49\textwidth}
    \caption{}
    \includegraphics{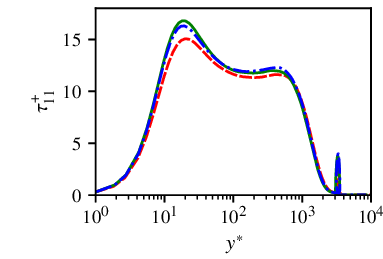}
    \label{fig:reystressx_st2}
\end{subfigure}
\caption{Wall-normal profiles of the density-scaled streamwise Reynolds stress for the
reactive (SBLI-R), thermally perfect (SBLI-TP), and calorically perfect
(SBLI-CP) simulations at (a) $\hat{x}=-4$ and (b) $\hat{x}=4$.}
\label{fig:reystress_x}
\end{figure}

Lastly, the wall-normal distribution of the density-scaled streamwise Reynolds stress is shown in figure~\ref{fig:reystress_x} for the upstream and downstream stations. Upstream of the interaction, all cases collapse closely, with only minor deviations in the outer layer. Downstream of the interaction, however, the reactive simulation exhibits the lowest peak levels, indicating that the thermodynamic modulation induced by chemistry also affects the turbulent velocity fluctuations.

Overall, the comparison reveals a clear hierarchy in the sensitivity of the present SBLI to fluid modelling. The difference between CP and TP descriptions remains secondary, whereas finite-rate chemistry produces systematic and measurable changes in the interaction. In particular, relative to the reactive case, simplified frozen models predict slightly larger separation bubbles, higher downstream wall heat flux, and stronger thermal fluctuation levels. 
For the present configuration, simplified fluid models therefore provide conservative estimates of wall heat flux and separation extent. However, this conservative character does not extend to all quantities: wall-pressure fluctuations are only weakly affected and do not exhibit the same monotonic ordering, with the CP case yielding slightly lower r.m.s. levels downstream of the interaction. This is an important practical outcome of the present comparison, since it indicates that simplified frozen models may still be useful for preliminary aero-thermal design when conservative estimates of specific observables are sought, even though finite-rate chemistry remains necessary for quantitative accuracy.

\section{Conclusions}
\label{sec:conclusions}
The present work has investigated the effects of thermochemical modelling on a turbulent hypersonic shock-wave/boundary-layer interaction by means of three direct numerical simulations performed with the same geometry, external flow conditions, and numerical methodology, but with progressively simplified fluid descriptions. The configuration considered corresponds to an oblique shock impinging on a fully turbulent high-enthalpy boundary layer at edge Mach number $M_e=6.385$, with edge stagnation enthalpy $H_e=16.9$ MJ/kg and shock-generator deflection angle $\beta=8^\circ$. The comparison between a finite-rate reactive case, a single-species thermally perfect gas model, and a single-species calorically perfect gas model was designed to isolate, as clearly as possible, the role of chemical non-equilibrium and thermal inertia in determining both the incoming boundary-layer state and the following SBLI response.

The analysis of the reactive simulation first shows that the shock system induces a marked thermochemical response of the boundary layer. Shock compression leads to a strong temperature rise, which is responsible for further activating chemical reactions and therefore significant modifications of the chemical composition after the interaction. However, the thermal and chemical responses are not synchronised. While temperature increases rapidly within the interaction region, peak chemical activity is shifted downstream, as indicated by the species wall-normal profiles, and the streamwise evolution of wall composition and Damk\"{o}hler numbers. The reacting mixture, therefore, does not adjust instantaneously to the thermal forcing imposed by the shocks but exhibits a finite-rate lag extending beyond the interaction length. In addition, the computed values of the turbulent Damk\"{o}hler numbers within the interaction region reach values of order unity for the most chemically active species, namely atomic and molecular oxygen. This suggests high levels of turbulence-chemistry interaction, which has direct implications in the modelling of the unclosed chemical source term in both LES and RANS formulations.

When the three fluid models are compared, a clear hierarchy emerges. The thermally perfect and calorically perfect models produce very similar predictions for most of the considered observables, indicating that within the class of single-temperature frozen-gas descriptions considered here, the choice of caloric closure has only a secondary influence on the interaction. By contrast, finite-rate chemistry yields significant changes in the mean thermal field, wall loads, separation topology, and reflected-shock angle. Relative to the frozen cases, the reactive simulation exhibits a slightly delayed separation and therefore a smaller separation bubble, lower post-interaction wall heat flux, lower mean and fluctuating temperature levels, and less inclined reflected shock due to variations in fluid properties. These trends are consistent with the additional energy sink associated with endothermic reactions, which absorbs part of the thermal energy induced by the shock and thereby weakens the overall interaction.
The results are also useful from a modelling and engineering standpoint. For the present case setup, simplified frozen models tend to over-predict both the wall heat flux and the extent of the separation. In this sense, they provide conservative predictions for two quantities that are central to the aero-thermal design of high-speed vehicles. This conservative character should not be generalised to all observables, however, since the calorically perfect gas assumption yields moderately lower pressure fluctuation levels downstream of the interaction. Simplified models may therefore remain useful for preliminary design estimates of selected thermal and separation quantities, but finite-rate chemistry is required for quantitative accuracy across the full set of observables.

More broadly, the present results suggest that, in the turbulent hypersonic SBLI considered in this study, the dominant modelling distinction is between frozen and chemically reacting descriptions, with the thermal modelling being only a secondary player. However, the conclusion drawn here should be interpreted in light of the specific interaction considered. Since stronger impinging shocks are expected to amplify both the separation extent and the thermochemical response, whether the same conclusions persist at larger interaction strengths remains an open question and deserves dedicated investigation.
A second open issue concerns the interplay between finite-rate chemistry and the low-frequency shock unsteadiness. The low-frequency to chemical time scale ratio shows that the dominant frequency of the breathing motion of the separated region and the time scales of the most active chemical species fall in the same range, suggesting the possibility of a non-trivial coupling. A more quantitative characterisation of this mechanism would require significantly longer sampling windows and is left for future work.

%\begin{bmhead}[Xxxxxxx.]
%For the custom heading such as acknowledgment, funding disclosure,
%conflict disclosure and any other like-wise sections must be
%mentioned in the optional braces as shown in this example.
%\end{bmhead}

\begin{bmhead}[Acknowledgments.]
We acknowledge that the results reported in this paper have been achieved using the EuroHPC JU Extreme Scale Access Infrastructure resource LUMI hosted at CSC's data center, Kajaani, Finland, under project EHPC-EXT-2024E02-048. We also acknowledge the CINECA award under the ICSC initiative (HPC grants CNHPC\_1662142 and CNHPC\_2021201), for the availability of high-performance computing resources on Leonardo booster. 
\end{bmhead}

\begin{bmhead}[Funding.]
We acknowledge financial support under the National Recovery and Resilience Plan (NRRP),
Mission 4, Component 2, Investment 1.1, Call for tender No. 104 published on 2.2.2022 by the Italian Ministry of University and Research (MUR), funded by the European Union – NextGenerationEU– Project Title ADMIRE - CUP B53C24006770006 - Grant Assignment Decree No. 1401 adopted on 18/09/2024 by the Italian Ministry of University and Research (MUR). This research received also financial support from ICSC – Centro Nazionale di Ricerca in “High Performance Computing, Big Data and Quantum Computing”, funded by European Union – NextGenerationEU.
\end{bmhead}

\begin{bmhead}[Data availability statement]
The data that support the findings of this study are available upon reasonable
request.
\end{bmhead}

\begin{bmhead}[Declaration of interests.]
The authors report no conflict of interest.
\end{bmhead}

\bibliographystyle{jfm}
\bibliography{bibl}

\end{document}